\documentclass[showpacs,aps,twocolumn]{revtex4}
\usepackage{graphicx}
\usepackage{amsmath,amssymb,amsfonts}
\usepackage{array}
\usepackage{url}
\usepackage{hyperref}
\usepackage{multirow}
\usepackage{float}
\usepackage{lineno}
\usepackage{xspace}
\usepackage{ulem}
\usepackage{natbib}
\usepackage{amsbsy}
\usepackage{amssymb}
\usepackage{amsmath}
\usepackage{graphicx}
\usepackage{silence}

\usepackage{array}
\usepackage{mathrsfs}
\usepackage{xfrac}
\usepackage{booktabs}
\usepackage{amsthm}
\usepackage{mathtools}
\usepackage{etoolbox}
\apptocmd{\sloppy}{\hbadness 10000\relax}{}{}
\usepackage{nameref}
\usepackage{hyperref}
\usepackage{url}      % Handles URLs properly in text and footnotes

\hypersetup{colorlinks=true, citecolor=blue, urlcolor=blue, linkcolor=blue}
\usepackage{xcolor}

\usepackage[english]{babel}
\newcommand{\beqa}{\begin {eqnarray}}
\newcommand{\eeqa}{\end {eqnarray}}

\def\be{\begin{equation}}
\def\ee{\end{equation}}

\def \nh{H{\sc I}}

\begin{document}

\title{Probing Decaying Dark Matter Using the  Post-reionization \nh ~21-cm signal}
\author{Mohit Yadav}
\email{p20210462@pilani.bits-pilani.ac.in}

\author{Tapomoy Guha Sarkar}
\email{tapomoy@pilani.bits-pilani.ac.in}

\affiliation{Department of Physics, Birla Institute of Technology and Science, Pilani, Pilani - 333031, Rajasthan, India}

\begin{abstract}
We propose the HI 21-cm  power spectrum from the post-reionization epoch as a probe of a cosmological model with decaying dark matter particles. The unstable  particles are assumed to undergo a 2-body decay into a massless and massive daughter. 
 We assume that a fraction $f$ of the total dark matter budget to be unstable and quantify the decay using the life time $\Gamma^{-1}$ and the relative mass splitting $\epsilon$ between the parent and the massive daughter. The redshift space anisotropic power spectrum of the post-reionization 21-cm signal brightness temperature, as a tracer of the dark matter clustering, imprints the decaying dark matter model through its effect on background evolution and the suppression of power on small scales. 
  We find that with an idealized futuristic intensity mapping experiment with a SKA1-Mid like radio-array, $\epsilon$ and $\Gamma$ can be measured at $3.37\%$ and $6.86\%$ around their fiducial values of $\epsilon = 0.012$   and $\Gamma = 0.008 {\rm Gyr}^{-1}$ respectively. When only the foreground-free window is considered in the $k-$ space,  these percentage errors in $\epsilon$ and $\Gamma$  degrade to  $9.91\%$ and $12.03\%$ 
 respectively.
 The forecasts under identical assumptions for a second, literature-anchored benchmark ($\epsilon=1.14\times10^{-4}$, $\Gamma=1.04\times10^{-3}\,\mathrm{Gyr}^{-1}$) yields projected uncertainties of $6.16\%$ on $\epsilon$ and $8.29\%$ on $\Gamma$ while, restricting to the foreground-free window increases these to $18.77\%$ and $14.22\%$, respectively.
\end{abstract}
\date{\today}
\maketitle

\section{Introduction}
Dark matter (DM) comprises roughly $\sim 25\%$ of the total energy density of the universe and plays the most crucial role in cosmological structure formation. Its existence is supported by a diverse range of observations, including galactic rotation curves \cite{brownstein2006galaxy,rubin1970rotation,botteon2019particle}, large-scale structure surveys \cite{zaldaloeb,Percival_2007, Okada_2013}, strong and weak gravitational lensing \cite{Jain_1997,Hui_1997,Seljak_1999,Bartelmann_2001,Huterer_2002,Hu_2002,waerbeke2003,Heavens_2003,Takada_2003,knox2004weak,Huterer_2006,Miyazaki_2007,MUNSHI_2008,Hoekstra_2008,Vallinotto_2009,Takada_2009,Smith_2009,GSarkar_2010,Kamionkowski_2016}, and anisotropies in the cosmic microwave background (CMB)\cite{sachs_wolfe,Komatsu_2009,Planck2018,k2004physics,hu2008lecture,tanaka2020detectability,Hu_2001,refId0}. In the standard cosmological framework, the $\Lambda$CDM model proposes that dark matter is cold, collisionless, and stable, offering a robust framework for understanding the expansion history of the universe and structure formation. Despite its remarkable success, the $\Lambda$CDM model faces observational challenges \cite{Perivolaropoulos_2022,1995crisis}. Notable amongst these issues is the Hubble tension \cite{hu2023hubble,abdalla2022cosmology} indicating a discrepancy between the  Hubble constant ($H_0$) measured from distance ladder estimates using Cepheid variables \cite{freedman2021measurements,lee2021astrophysical,freedman2023progress, sandage2006hubble,beaton2016carnegie,Riess_2016} and CMBR observations \cite{PhysRevD.104.083509,jain2003cross,huterer2002weak,martinet2021probing,amendola2008measuring}.
There is also the consistent mild tension $\sim 2-3\sigma$ in the clustering amplitude quantified through $S_8 = \sigma_8 ( \Omega_m/0.3)^{0.5}$ measured CMB and weak lensing surveys, with the latter consistently reporting lower values of $S_8$ \cite{Planck1(2020),Kukijen(2019),Giblin(2021),Wright(2021),Aihara_2017,Hamana(2020),Dark_Energy_Survey(2005),Amon(2022)}. 
These cosmological tensions seem to have often motivated investigation of  physics beyond the standard $\Lambda$CDM paradigm, and attempts have been made to modify both the  dark energy sectors \cite{Perivolaropoulos_2022,Ruchika(2020),Visinelli(2019),Guo(2019),Yadav(2019),Poulin(2019),Niedermann(2019),Sakstein(2020),Valentino(2018),Banihashemi1(2019),Banihashemi2(2020),Mena(2020),Supriya1(2019),Supriya2(2019),Supriya3(2018),Dhawan(2018),Alestas(2020),Yanagida(2020)} and dark matter sectors \cite{Raveri(2017),Nagata(2017),Silva(2019),Emami(2018),Alcaniz(2021),Eramo(2018),Loeb(2019),Goldberg(2015),Vikram(2017),Torsten(2018),Pandey(2020),Klypin(1988),Hopper(2012),Zant(2019),Nima(2019),Renk(2017),Mario(2020),Mario2(2020)}.
Any modification of the pre-recombination physics aims to decrease the sound horizon at recombination to alleviate the Hubble tension. However, this has the possibility of spoiling the bounds on other parameters and may even aggravate the $S_8$ tension.
To resolve the $S_8$ tension, it is typically required to reduce the amplitude of matter fluctuations on small scales. Such suppression can be achieved through modifications of the  dark matter properties at low redshifts \cite{sCHNEIDER(2020),Parimbelli(2021),Duccio(2016),Heimersheim(2020),Racine(2016),Zavala(2016),Rubira(2022),Joseph(2023),Skordis(2013),
2023PhRvD.107l3538P}. Unstable dark  matter  with a massless relativistic daughter can potentially cause such a suppression owing to the erasure of power on small scales \cite{Kari(2015),Kari2(2020),Murgia(2017),Abell(2021),Chen(2021),Choi(2022),Bucko(2023)}

There is no strong argument from fundamental physics \cite{Bomark(2014),Ibarra(2009),Asaka(2007),Ibarra(2013),Hooper(2007)} or observations \cite{Aartsen(2014),Accardo(2014),Aguilar(2014),Anchordoqui(2014)}  to believe that dark matter is entirely stable.   Decaying Dark Matter (DDM) models, which permit dark matter to decay into lighter particles such as dark radiation and warm dark matter, represent a natural extension of the standard model. Such decays alter the evolving energy density of the universe, consequently affecting its expansion history and inhibiting structure formation on small scales. These effects make DDM particularly important for addressing the persistent cosmological tensions that challenge the $\Lambda$CDM model. Through the redistribution of energy between radiation and matter, DDM possesses the capacity to mitigate these tensions while adhering to current cosmological data. Moreover, DDM scenarios may also resolve issues with small-scale structures, including "cuspy core" \cite{de2010core} and "missing satellite" problem \cite{Bullock(2010)}, by reduction of clustering on small scales.

The dynamics of DDM is defined by two principal parameters: the decay rate ($\Gamma$), which determines the lifetime of the particle, and the fraction of rest mass energy ($\epsilon$) allocated to the decay products\cite{ibarra2013indirect,abellan2021linear,Vattis,blackadder2016cosmological,vattis2019dark,clark2021cosmological,Giri(2024)}. These parameters imprint on the background evolution, and development of cosmological perturbations, thereby influencing observables like the matter power spectrum. Given our interest in late-time cosmology, we are interested in the tomographic study of the post-reionization universe using the redshifted HI 21-cm signal. 

The post-reionization IGM for $z <6$ is predominantly ionized. While this is true for the diffuse low density gas, a small fraction of neutral
hydrogen survives in  confined over-dense regions and is self-shielded from the background ionizing radiation.
These clumped, dense self-shielded, damped Lyman-{$\alpha$} systems (DLAs) \cite{wolfe05} are located in regions of highly non-linear matter over density peaks \cite{coo06, zwaan, nagamine}, and are believed to store bulk ($\sim 80\%$) of the \nh  ~ at $z<4$ \cite{proch05} with \nh ~
column density greater than $ 2 \times 10^{20}$atoms/$\rm cm^2$
\cite{xhibar, xhibar1, xhibar2}. These DLA clouds, source the 21-cm radiation emission in the post
reionization epoch.
The post-reionization  HI 21-cm signal has been extensively studied \cite{ poreion0, poreion1, poreion2, poreion3, poreion4, poreion5, poreion6,  poreion8, param2}
and is a powerful cosmological probe  \cite{param1, Bull_2015, param2, param3, param4}. 

The first detection of intensity mapping of 21-cm signal in the redshift range $z \sim 0.53 - 1.12$ was achieved using radio observations with the Green Bank Telescope (GBT) \cite{chang2010hydrogen} indicating that the 21-cm intensity field traces the distribution of galaxies at $z \approx 1$ \cite{davis2003deep2}. The 21-cm signal was further detected in  cross-correlation between the 21-cm signal and galaxies in the WiggleZ Dark Energy Survey \cite{masui2013measurement}  at $z \approx 0.8$. An observational upper limit on the 21 cm auto-power spectrum has also been reported \cite{switzer2013determination}. The 21-cm power spectrum in auto-correlation has yet to be discovered. This is a primary science goal of several presently operational as well as new or upgraded radio interferometric arrays. Some of these telescopes which aim to detect the cosmological 21cm signal are:  the Giant Metrewave Radio Telescope (GMRT)\footnote{https://www.gmrt.ncra.tifr.res.in}, the Ooty Wide Field Array (OWFA)\footnote{http://rac.ncra.tifr.res.in/ort/owfa.html}, The Canadian Hydrogen Intensity Mapping Experiment (CHIME)\footnote{https://chime-experiment.ca/en}, the Meer-Karoo Array Telescope (MeerKAT)\footnote{https://www.sarao.ac.za/science/meerkat/about-meerkat/}, the Square Kilometer Array (SKA)\footnote{https://www.skao.int/en}, HIRAX \footnote{ https://hirax.ukzn.ac.za/} .
 A detection of 21 cm emission from large-scale structure (LSS) between $z =0.78$ and $z= 1.43$ using the cross-correlations of CHIME data with eBOSS catalogs of luminous red galaxies (LRGs), emission-line galaxies (ELGs), and quasars (QSOs) has been recently reported \cite{amiri2023, amiri2024}. 
  A detection of the correlated clustering between MeerKAT intensity maps and galaxies from the WiggleZ Dark Energy Survey has also been recently reported at a $7.7\sigma $  level\cite{meerkat1}.  
These promising observations motivate intensity mapping science goals of telescopes like SKA, GMRT, and  HIRAX.

The parameters of a model involving decaying dark matter shall affect the cosmological background evolution and matter clustering. In this paper, we study the post-reionization HI 21-cm anisotropic power spectrum in redshift space in a cosmology with decaying dark matter and make error projections of model parameters using a futuristic intensity mapping experiment with a SKA-Mid-like radio interferometer.

\section{Cosmology with decaying dark matter (DDM)}
In the simplest case, an unstable dark matter particle decays into a single relativistic component known as dark radiation. Such models are very well constrained from CMBR observations \cite{Hubert(2021),Simon(2022),Jozef(2023)}.
We shall consider a 2-body decay scenario \cite{Simon(2022),Abell(2021),Wang(2012),Annika(2010),Cheng(2015),Mau(2022)} where a fraction $f$ of the dark matter component consists of unstable particles (denoted by the subscript $"0"$)   decays into a massless dark photon (denoted by the subscript $"1"$) \cite{Rajaraman(2003),Kopp(2018)}  and massive daughter particles  (denoted by the subscript $"2"$) with mass $m_2$.
The parent unstable particle of mass $m_0$ assumed to be at rest,  decays exponentially with a lifetime $\tau = 1/\Gamma$.
If $\epsilon$ denotes the fraction of energy of the parent particle transferred to the massless daughter, then from conservation of energy and momentum, 
\begin{equation}
\epsilon = \frac{1}{2} \left ( 1 - \frac{m_0^2}{m_2^2} \right ) ~~~~~~\&~~~~~~\beta_2 = \frac{v_2}{c} = \frac{\epsilon}{1 - \epsilon}
\end{equation}
where $v_2$ is the magnitude of the "kick" velocity of the massive daughter. In the non-relativistic limit, $v_2 = \epsilon c$. 
Thus, the dark-matter sector now comprises four species of particles: Stable cold dark matter (CDM), parent unstable decaying dark matter (DDM), massive daughter and massless dark photon. 
The evolution of the densities $(\rho_{\rm cdm} , \rho_0, \rho_1, \rho_2)$ of these particles in an expanding universe is governed by \cite{Vattis,blackadder2016cosmological,vattis2019dark,clark2021cosmological,Giri(2024),Blackadder2014}
\begin{eqnarray}
\frac{d \rho_{\rm cdm}}{d t} &+&3 \frac{\dot{a}}{a} \rho_{\rm cdm} = 0 \\
\frac{d \rho_0}{d t} &+& 3 \frac{\dot{a}}{a} \rho_0 = -\Gamma \rho_0 \\
\frac{d \rho_1}{d t} &+&4 \frac{\dot{a}}{a} \rho_1 =\epsilon \Gamma \rho_0 \\
\frac{d \rho_2}{d t} &+&3\left(1+w_2(a)\right) \frac{\dot{a}}{a} \rho_2 = (1-\epsilon) \Gamma \rho_0
\end{eqnarray}
where $w_2(a)$ is the equation of state for the massive daughter particle.
The first two equations are not coupled and can be solved with suitable initial conditions, as
\begin{equation}
\rho_{\rm cdm} (a) = ( 1 -f ) \frac{\Omega_{mo}}{a^3}~~\&~~\rho_0(a) = \mathcal{A} \frac{e^{-\Gamma t(a)}}{a^3},
\label{eq:rho0}
\end{equation} 
where $\mathcal{A}$  is a normalization set by the initial condition on
$\rho_0(a)$. Assuming that dark matter does not decay prior to recombination ($z_{\rm rec} \sim 1090$), this normalization is obtained by the condition $\rho(a_{\rm rec}) = f \rho_{\rm crit} \Omega_{mo} ( 1 + z_{\rm rec})^{3}$, where $\rho_{\rm crit} = 3H_0^2/8\pi G$, and $\Omega_{mo}$ is the total dark matter density parameter at the present epoch.
\begin{figure}[ht!]
    \centering
    \includegraphics[width=0.40\textwidth]{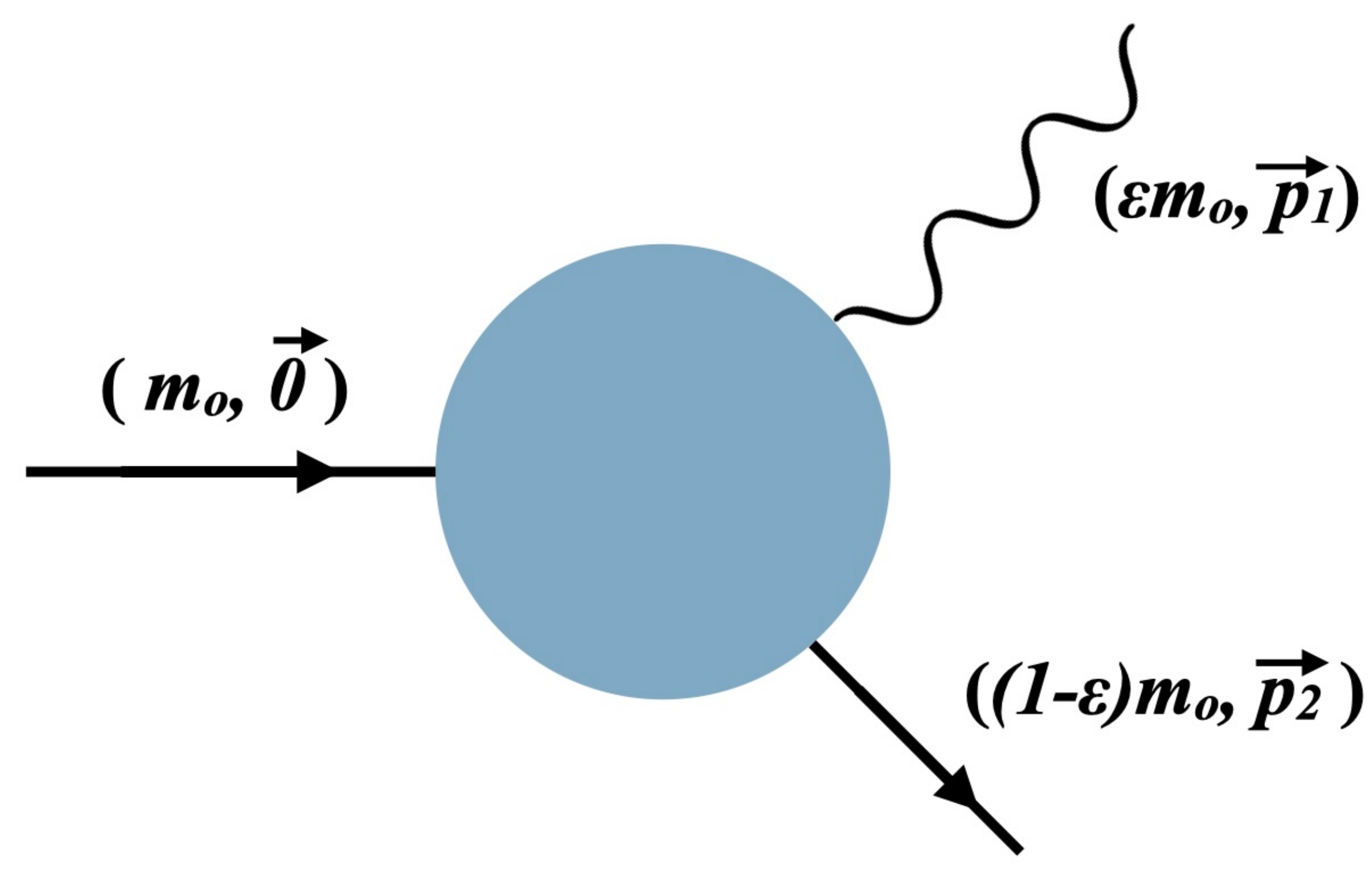}
    \caption{A schematic representation of the 4-momentum $p^{\mu} =(E, {\mathbf p})$ in a two-body decay with a massive parent particle decaying into a massless and  massive daughter}
    \label{fig:1}
\end{figure}

Using the solution $\rho_0(a)$ from Eq. {\ref{eq:rho0}} we have  
\begin{equation}
d\left(\rho_1 a^4\right)=\epsilon  \mathcal{A}  \Gamma  e^{-\Gamma t} a dt 
\end{equation}
On integration, this gives us 
\begin{equation}
\rho_1(a)=\frac{\epsilon \mathcal{A}}{a^4}\left(\int_{a_{\rm rec}}^a e^{-\Gamma t\left(a^{\prime}\right) d a^{\prime}}-a e^{-\Gamma t(a)}+a_{\rm rec} e^{-\Gamma t\left(a_{\rm rec}\right)}\right)
\end{equation}

The evolution of the massive daughter is complicated since its equation of state $w_2(a)$ is dynamic. There is a possibility for this particle to be relativistic initially and slow down later. If this transition from a relativistic to a non-relativistic regime occurs at $a_D$ for a given particle, then one has to consider the sum of contributions to $\rho_2(a)$ from different particles. 
Thus we have \cite{Blackadder2014}

\begin{equation}
\rho_2(a)= \frac{\mathcal{A} \Gamma \sqrt{1-2 \epsilon}}{a^3} \int_{a_*}^a \mathcal{K}\left(a, a_D\right) d a_D
\label{eq:rho2}
\end{equation}

where

\begin{equation}
\mathcal{K}\left(a, a_D\right) \equiv \frac{e^{-\Gamma t\left(a_D\right)}}{a_D H_D} \sqrt{\frac{\beta_2^2}{1-\beta_2^2}\left(\frac{a_D}{a}\right)^2+1}
\end{equation}

We note that to evaluate $\rho_0, \rho_1,  \rho_2$ ,we require $t(a)$ and $H(a)$ which are interdependent on each other, while $H(a)$ itself depends on $\rho_0, \rho_1,  \rho_2$.
Thus, these equations are solved iteratively to obtain $H(a)$.

\begin{figure}[ht!]
    \centering
    \includegraphics[width=0.45\textwidth]{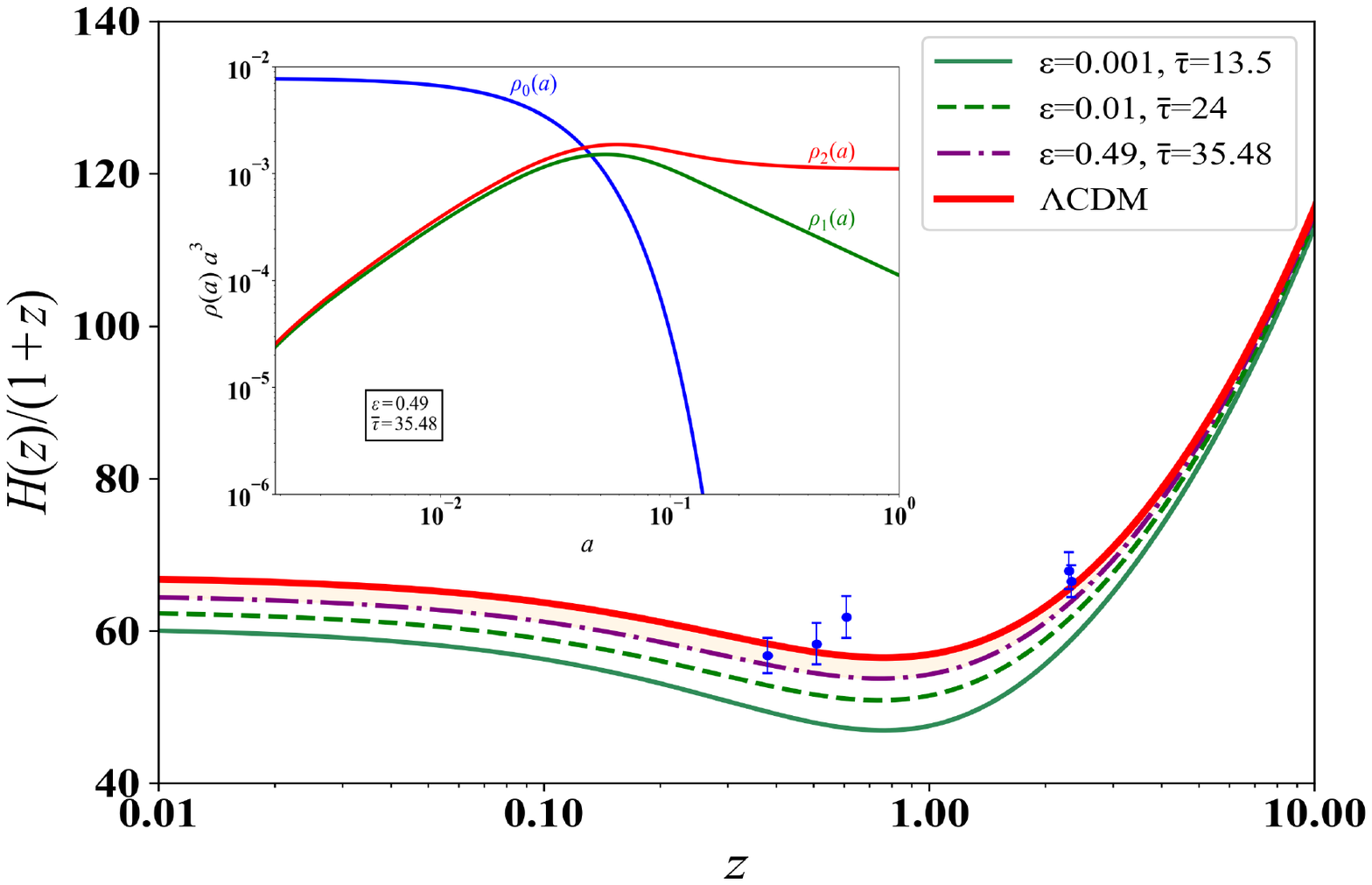}
    \caption{The Hubble parameter $H(z)$ for different $(\bar{\tau} = \tfrac{\Gamma^{-1}}{1 {\rm Gyr}}, \epsilon)$ shows the expected match with $\Lambda$CDM model at high redshifts and departure at low redshifts. The inset shows the evolution of the parent DDM and daughter particles. Here, we have assumed $f=1$.The observational results at $z=0.38$, $0.51$, and $0.61$ are from galaxy clustering data of BOSS SDSS III \cite{sdss3}. We have also shown some high redshift data from Lyman-$\alpha$ forest at $z=2.3$ \cite{boss-lydata2.3} and $z = 2.4$ \cite{boss-lydata2.4}.}
    \label{fig:hubbleplot}
\end{figure}

\begin{equation}
H^2(a) = \frac{H^2_0}{\rho_{\rm crit}} \sum \rho_i(a)
\label{eq:hubble}
\end{equation}
where we have CDM, baryonic matter, radiation and $\Lambda$  contributing to the sum other than the contributions from DDM and its daughter particles.

Fig. \ref{fig:hubbleplot}, shows the expansion history for different DDM parameters. The decay of DDM starts at $z = 1090$. Thus, for sufficiently small $\epsilon$ and large $\tau$, the Hubble parameter follows the $\Lambda$CDM behaviour at high redshifts.
Only at low redshifts and much after the epoch of recombination does the DDM 
density falls, while the daughter particles are produced as seen in the inset of Fig. \ref{fig:hubbleplot}. The lifetime $\Gamma^{-1}$, thus controls the redshift when the expansion history deviates from  $\Lambda$CDM whereas $\epsilon$ determines the fraction of DDM density that goes into the daughters.

In the two-body DDM model, the parameters have clear physical roles. The mass--energy fraction \(\epsilon\) transferred to the daughter sets the daughter particle's kick velocity,
\[
\frac{v_{\rm kick}}{c}=\frac{\epsilon}{1-\epsilon}\simeq \epsilon \quad (\epsilon\ll 1),
\]
which determines the free-streaming scale and thus the scale dependence of the small-scale power suppression. The decay rate \(\Gamma\) defines the lifetime \(\tau\equiv \Gamma^{-1}\) and controls when the suppression accumulates, i.e.\ the late-time amplitude and its redshift evolution. We adopt \(f=1\) (all DM decays), the standard two-body scenario in recent global analyses, which also maximizes detectability in a forecast; results for \(f<1\) follow by an approximate rescaling of the suppression amplitude.

At low redshifts, the Hubble rate falls below $\Lambda$CDM, owing to an excess of radiation. It must be noted that we have kept the dark energy sector fixed, $\it i.e.$ $\Omega_{\Lambda} = 0.69$ is held fixed. However, if we impose an additional constraint at very high redshift that the angular diameter distance to the epoch of recombination is strongly constrained from observations,  then it would require $\Omega_{\Lambda}$ to be strongly correlated with $\epsilon$ and $\Gamma$, whereby the necessary increase in $\Omega_{\Lambda}$ will actually make $H(z)$ larger than the $\Lambda$CDM expectation in the dark energy dominated era.

A crucial cosmological impact of two-body decaying dark-matter scenarios is that small-scale structures are wiped out. This leads to a suppression of the linear matter power
spectrum at scales smaller than a 'cut off scale' $k > k_f$.
The two parameters of the DDM model $(\epsilon, \Gamma) $ have two distinct roles to play in this suppression.
The parameter $\epsilon$, which measures the fraction of the mass of the DDM particle that transformed into kinetic energy, 
fixes the value of $k_f$. The higher the energy, the smaller the cutoff $k_f$.
If the lifetime of DDM $\tau = \Gamma^{-1}$  is small, the abundance of the free-streaming decay products shall be large.
Thus, $\Gamma$
determines the amount of small-scale power suppression. In addition, the effects
of decay become more pronounced at low redshifts for a given $\Gamma$,
due to the greater abundance of the daughter particles at low redshifts. Thus, the power suppression is more at lower redshifts. 
All these effects are incorporated into the transfer function of the matter power spectrum at $ z < 2.35$ \cite{Potter(1),Peters(1)}.

\section{The post-reionization HI 21-cm power spectrum}

 The IGM is mostly ionized at redshifts $z \leq 6$. In this epoch, the 21-cm signal, which is seen in emission, is predominantly sourced by DLA clouds that remain neutral as they are self-shielded from the background ionizing photons. 
In an intensity mapping experiment, one is typically interested in mapping the collective diffuse emission from these gas clouds.  The low-resolution large-field imaging, thus, ignores the discrete nature of DLA sources and the associated Poisson noise. This is a reasonable assumption, since the number density of
the DLA emitters is very large \cite{poreion8}. The actual distribution of HI in the post-reionization epoch is not known to us. 
The modeling of the post-reionization \nh ~ 21-cm signal is, thus based on implicit observations, or is derived from numerical simulations.

At low redshifts, the population of the triplet state of \nh, is enhances by Wouthuysen field effect  ~causing the ratio of the spin temperature and the CMB temperature $T_s/T_{\gamma}$ to be much larger than unity. As a result, the 21-cm radiation appears as an emission signal against the CMBR \cite{madau97, bharad04, zaldaloeb}.
There is strong observational evidnce from the Lyman-$\alpha$ absorption lines in quasar spectra that, in the post-reionization epoch the mean neutral fraction
 $\bar{x}_{\rm HI} = \Omega_{gas}/\Omega_b \sim 2.45\times 10^{-2}$,
which does not indicate any evolution in the redshift range $ z \leq 6$ \cite{proch05}.

The  \nh ~in the post-reionization epoch is predominantly ($\sim 80\%$) in the  DLAs, which are believed to be galaxies.  Galaxies are well-known dark matter tracers \cite{dekel, mo, yosh}. Thus the \nh~distribution is expected to trace the underlying dark matter density field  with a possible scale and redshift dependent bias $b_T(k, z)$ defined as ${b}_T(k, z) = {\left [\frac{P_{\rm HI}(k, z)}{P(k, z)}\right ]}^{1/2}
$where  ${P}_{\rm HI}(k, z)$ and ${P}(k, z)$ denote the \nh ~and  dark matter power spectra, respectively. 
The dark matter distribution is assumed to be generated by a Gaussian random process. The statistics of such a Gaussian random field are thus entirely quantified using the 2-point function/power spectrum. 

 On large scales, the bias is known to be linear and scale-independent, but on small scales it is $k-$dependent. The post-reionization  \nh ~bias has been  studied  using N-body simulations \cite{bagla20, Guha_Sarkar_2012, Sarkar_2016, Carucci_2017}.
In these $N-$body simulations of the post-reionization \nh,  firstly, the dark matter distribution is simulated. Subsequently, dark matter halos are identified and populated with \nh~using some rule related to the mass of the halos.
 
These simulations generally indicate that the large-scale linear bias grows monotonically with redshift for $1< z< 4$ \cite {Mar_n_2010}. This feature is also seen in galaxy bias \cite{fry, mo, moo}.
There is a steep rise in the 21-cm bias on small scales. This is due to the absence of small mass halos as is expected from the CDM power spectrum and, consequently, the \nh ~being distributed only in larger halos. We have adopted the fitting formula for the bias $b_T(k, z)$ as a function of both redshift $z$ and
scale $k$ obtained from numerical simulations \cite{Sarkar_2016} of the post-reionization  signal. 
\be
\label{eqn:bias}
b_{T}(k,z) = \sum_{m=0}^{4} \sum_{n=0}^{2} c(m,n) k^{m}z^{n}
\ee 
{ The adopted expression for bias is obtained from cold dark matter simulations.
We point out here that the \nh~  bias needs to be obtained from DDM simulations. While we use the \nh ~bias in Eq. \ref{eqn:bias}, we note that the bias shall still remain linear on large scales even in DDM scenarios. 
On small scales, we know that DDM models shall have suppression of the dark matter power.
This would lead to a lesser abundance of low-mass halos. Subsequently, the gas would be distributed in larger halos. This would lead to an enhancement of the small-scale bias as compared to the CDM case.
Thus, our adoption of the bias in Eq. \ref{eqn:bias} is an underestimate. 
We also note that in our final analysis, we have marginalized over the bias parameters. This, however, leads to a massive increase in the errors.}

Adopting all the assumptions discussed above,  the power spectrum of post-reionization \nh~ 21-cm brightness temperature fluctuations from redshift $z$
is  given by \cite{bharad04, param3} 
\begin{widetext}
\be
\label{eq:21cmps}
P_{\rm HI }(k, z, \mu) = \frac{1}{\alpha_\parallel \alpha^2_\perp} \mathcal{C}^2(z) \left[ 1 + f_g \widetilde{\mu}^2 / b_T \right]^2 G_{_{FoG}}(k_{\parallel})  P\left( \widetilde{k}, z \right)
 \ee
 
Here $P(k, z)$ is the matter power spectrum at redshift $z$ and the  normalization $\mathcal{C}(z)$ is  given by 
\be
\frac{{\mathcal{C}(z)}}{4 \rm mK} = b_T(k,z) \bar{x}_{\mathrm{HI}}(1+z)^2\left(\frac{\Omega_{b 0} h^2}{0.02}\right)\left(\frac{0.7}{h}\right) \frac{H_0}{H(z)}, ~~{\rm and} \ee
\begin{equation}
\alpha_{\parallel} = \dfrac{H^{f}}{H^{r}},~~
\alpha_{\perp}   = \dfrac{D_A^{r}(z)}{D_A^{f}(z)},~~ 
\mu = \dfrac{k_{\parallel}}{k},~~
\widetilde{\mu}^{2} = \dfrac{\mu^{2}}{F^{2}+\mu^{2}(1-F^{2})}, ~~
\widetilde{k}  = \dfrac{k}{\alpha_{\perp}}
                      \sqrt{1+\mu^{2}\!\bigl(F^{-2}-1\bigr)},~~~
 F  = \dfrac{\alpha_{\parallel}}{\alpha_{\perp}}
\end{equation}
\end{widetext}
with $\mu
={\bf{\hat{k}}}\cdot{\bf{\hat{n}}}$, and $\beta_T(k, z) = f_g(k, z)/b_T(k, z)$, and where $f_g$ is the logarithmic growth rate of density fluctuations. 
Thus $f_g = d \ln D_{+}(a)/ d \ln a$, where the growing mode $D_+$ satisfies the equation  
\be \ddot D_{+} + 2 H \dot D_{+} - 4 \pi G \rho_M D_{+} = 0 \ee
where, $\rho_M = \rho_o 
+ (1 - 3 w_2) \rho_2 $.
The quantity $\mathcal{C}(z)$ which sets the amplitude of the power spectrum is 
thus determined by the mean neutral fraction $\bar{x}_{\mathrm{HI}}$, and the bias $b_T(k,z)$, and is  also sensitive to the evolution history $H(z)$.

The term $ f_g(z, k)  \mu^2$ has its origin in the  \nh ~peculiar
velocities \cite{poreion2, bharad04} which, as we mentioned, is sourced by the dark matter fluctuations.
On the large cosmological scales of interest, HI peculiar velocities are assumed to be determined by the dark
matter distribution. Thus, peculiar velocity manifests itself as a redshift space
distortion anisotropy (RSD) in the 21-cm power spectrum \cite{kaiser1987clustering,hamilton1998linear}). Other than this usual anisotropy of the power spectrum due to {\it redshift space distortion}, we have also incorporated the nonlinear effect of peculiar velocity dispersion - {\it Finger of God effect} \cite{jackson1972critique} and the { \it Alcock-Paczynski (AP) effect} \cite{AP1979, lopez2014alcock}. The Finger of God anisotropy is taken to be of the form
$ G_{_{FoG}}\left(k_{\|}, \sigma_p\right) = \left(1+\frac{1}{2} k_{\|}^2 \sigma_p^2\right)^{-2}$, where $\sigma_p$ is the pair velocity dispersion.
The AP effect manifests as geometric distortion of cosmological structures and arises due to departure of cosmological distances as predicted in the fiducial cosmology from their actual values along the line of sight and in the transverse direction.

 The DDM parameters $(\epsilon, \Gamma,  f)$  imprint on the  the 21-cm power spectrum through {\it (a) } the expansion rate $H(z)$ which affects the amplitude $\mathcal{C}(z)$  and {\it (b)} the small scale suppression of matter power spectrum $P(k,z)$  and {\it (c)} the anisotropy terms which are determined by $H(z)$, $D_A(z)$ and $f_g(z)$.

For a given set of the parameters $(\epsilon, \Gamma,  f)$, 
 $H(z)$ and subsequently $D_A(z)$ are obtained by numerically solving Eq. 2.6 - 2.10.
 The matter power spectrum $P(k,z)$ is obtained at the given redshift as the output of the emulator \cite{Giri(2024)}. 
 
We have taken the fiducial values of the parameters $(\Omega_m, \epsilon, \Gamma, H_0, f) = (0.3094, ~0.012,  ~0.008 {\rm Gyr^{-1}}, ~67.73 {\rm Km/s/Mpc}, ~1.0)$ from the best fit values from BestFit2 + KiDS \(S_8\) obtained in an earlier work\cite{ddmdata}.

These fiducial values correspond to a small mass splitting at the \(\sim 1\%\) level. This gives the massive daughter particle a mildly warm--DM--like velocity (\(v_{\rm kick}\simeq 0.01c\)), which suppresses structure below a characteristic free--streaming scale. The lifetime of \(\tau\sim 10^2\)~Gyr is much greater than the age of the Universe, meaning the parent particle is ``mostly stable'' but still has some late time imprint.  
Such a large lifetime also ensures that the background expansion history, \(H(z)\) and \(D_A(z)\), remain nearly identical to \(\Lambda\)CDM behaviour.  The observable impact is instead a redshift--dependent, scale--dependent suppression of structure growth which grows stronger toward lower redshift as more decays occur. This $ z <1$   regime is phenomenologically interesting because it can mildly alleviate the $S_8$ tension between CMB--inferred and low--redshift clustering amplitudes.

While our primary motivation to use the  chosen fiducial values rests of the data analysis  in  \cite{ddmdata},  we also note that Abellán, Murgia and Poulin (2021)~\cite{abellan2021linear} identify an \(S_8\)-alleviating ``ridge'' along which late--time, scale--dependent suppression fits current data while keeping background changes modest:
\[
\tau \simeq 55 \left(\frac{\epsilon}{0.007}\right)^{1.4}\ {\rm Gyr}.
\]
For our chosen \(\epsilon=0.012\), this relation gives \(\tau \simeq 1.17\times 10^2\,{\rm Gyr}\), which is of almost same order of magnitude with our fiducial \(\tau=1.25\times 10^2\,{\rm Gyr}\).
Thus, the Lyman--\(\alpha\) forest constraints by Fuß and Garny (2023)~\cite{ddmdata} used by us are consistent with the  parameter range predicted by \cite{abellan2021linear}.
Further, other independent observational bounds like EFTofLSS work (Simon et al.\ 2022)~\cite{Simon(2022)} combining Planck\,+\,EFTofBOSS\,+\,an \(S_8\) prior likewise prefers the same neighborhood (\(\epsilon\simeq 1\%\), \(\tau\sim 10^2\,\mathrm{Gyr}\)).

Recent non-linear modeling and emulator efforts have also focused on this range in the parameter space \cite{Giri(2024)}.
Hence, our fiducial model targets a phenomenologically interesting and currently viable region where 21\,cm IM  can be  most informative.

We note that, for the fiducial values used (especially $\Gamma$), the changes in $H(z)$ are very small. For a few percent change in $H(z)$ as in Fig. 2, we need $\tau = \Gamma^{-1}$ to  be almost a factor of $\sim 10$ times smaller. Such a behaviour is also reported in earlier studies\cite{abellan2021linear}. This implies that the predominant effect at these fiducial values is not on the background evolution, but the suppression of small-scale power.

For a given $\Gamma$, increasing \(\epsilon\) moves the cutoff scale for suppression to larger scales, while  decreasing \(\epsilon\) pushes suppression cutoff to noisy, smaller scales
For a given $\epsilon$, shorter lifetimes (larger \(\Gamma\)) enhance the late-time suppression amplitude and its redshift dependence.  

We note that the crucial effects come from non-linear scales $k_{\perp}, k_{\parallel} > 0.1 h/Mpc$, which implies other DM model variations and astrophysics can be important sources of confusion when interpreting our results. 

We consider three fiducial observational redshifts $z= 1.75,2.00,2.35$
and study some of these dark matter / astrophysical scenarios which maybe degenerate with the DDM models for the given fiducial values of the parameters. 

One-body decaying dark matter is a minimal extension of the $\Lambda$CDM model. In these models, dark matter decays into radiation \cite{Bertone,Feng}. 
The one-body decaying models where the dark matter is reasonably stable has power spectrum suppression of similar magnitude as in the two body decaying scenarios \cite{Bucko(2023)}. We consider one-body Decaying Dark Matter by making use of a publicly available emulator OBDemu\cite{Bucko(2023)}. 

Another modification to the standard cold dark matter paradigm is to consider warm dark matter. 
Warm dark matter particles are relativistic at early times and their density fluctuations are suppressed owing to free streaming on the scales
that are comparable to the horizon size at those epochs. This consequently leads to a 
suppression of the small-scale power.
 We adopt the standard {\it thermal relic} parameterization of the linear WDM transfer function from \cite{Vielwdm,Eisensteinwdm}, with $\mu=1.12$ and the $\alpha(m_{\rm WDM},\Omega_{\rm WDM},h)$ mapping. The WDM response is applied as a multiplicative, isotropic factor
$S_{\rm WDM}(k)\equiv[T_{\rm WDM}(k)/T_{\Lambda{\rm CDM}}(k)]^{2}$
to our fiducial redshift--space $P_{21}^{\Lambda{\rm CDM}}(k_\perp,k_\parallel,z)$.  We consider three typical  masses, $m_{\rm WDM}=\{0.5,0.3,0.2\}\,$keV, spanning mild to strong small--scale cut--offs.

\begin{figure}[h]
 \centering
\includegraphics[width=0.45\textwidth]{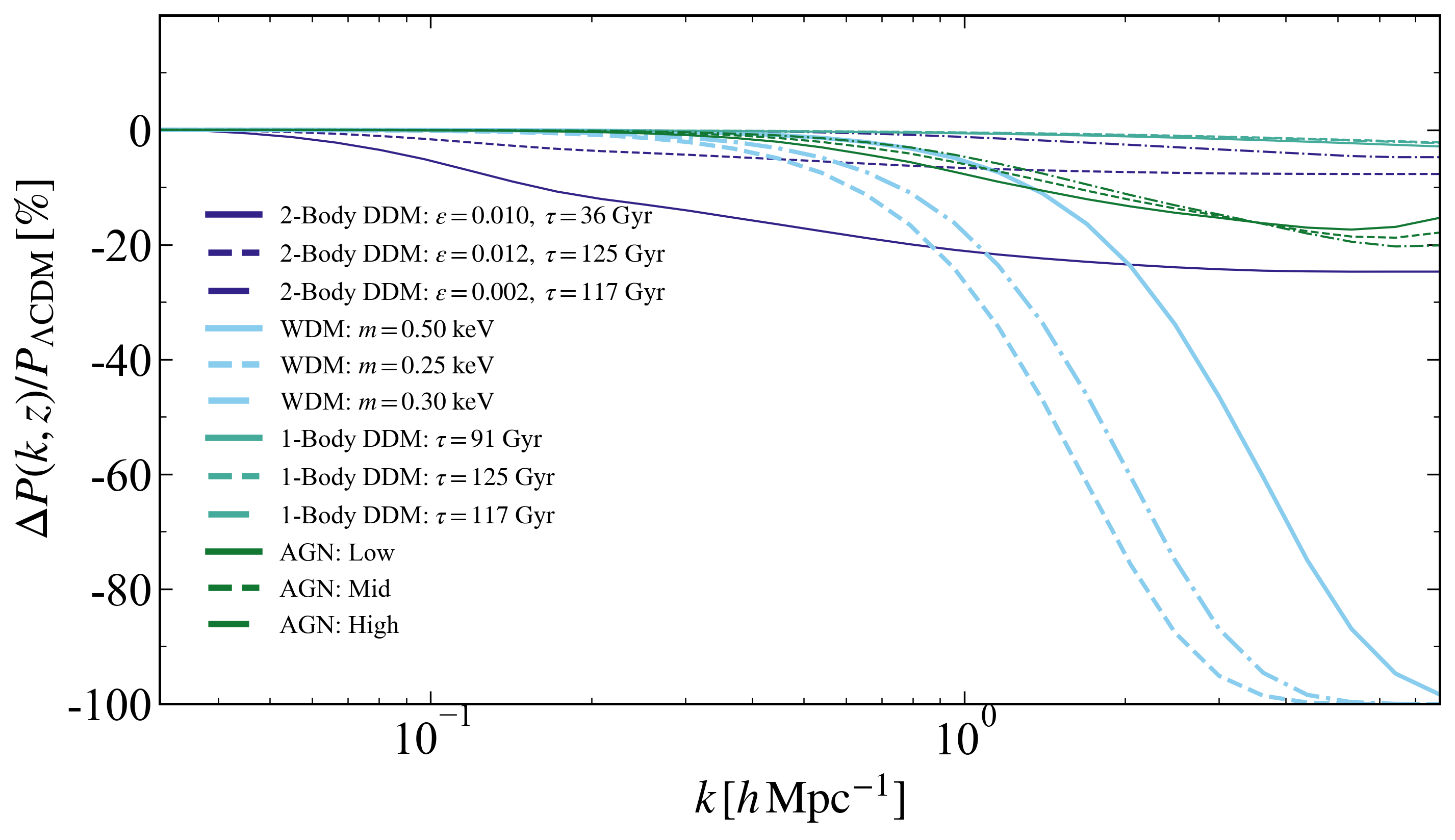}
\includegraphics[width=0.45\textwidth]{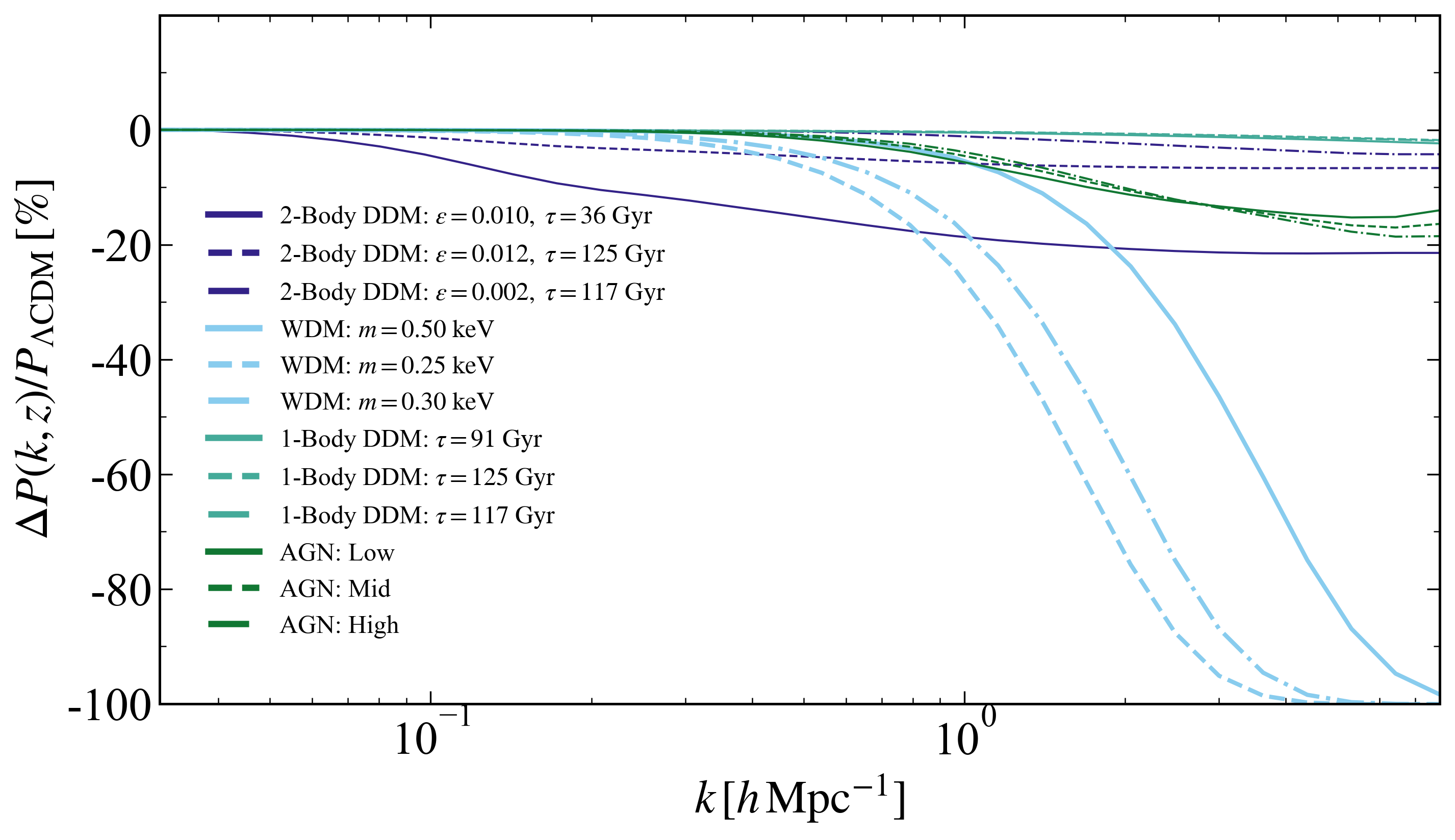}
\includegraphics[width=0.45\textwidth]{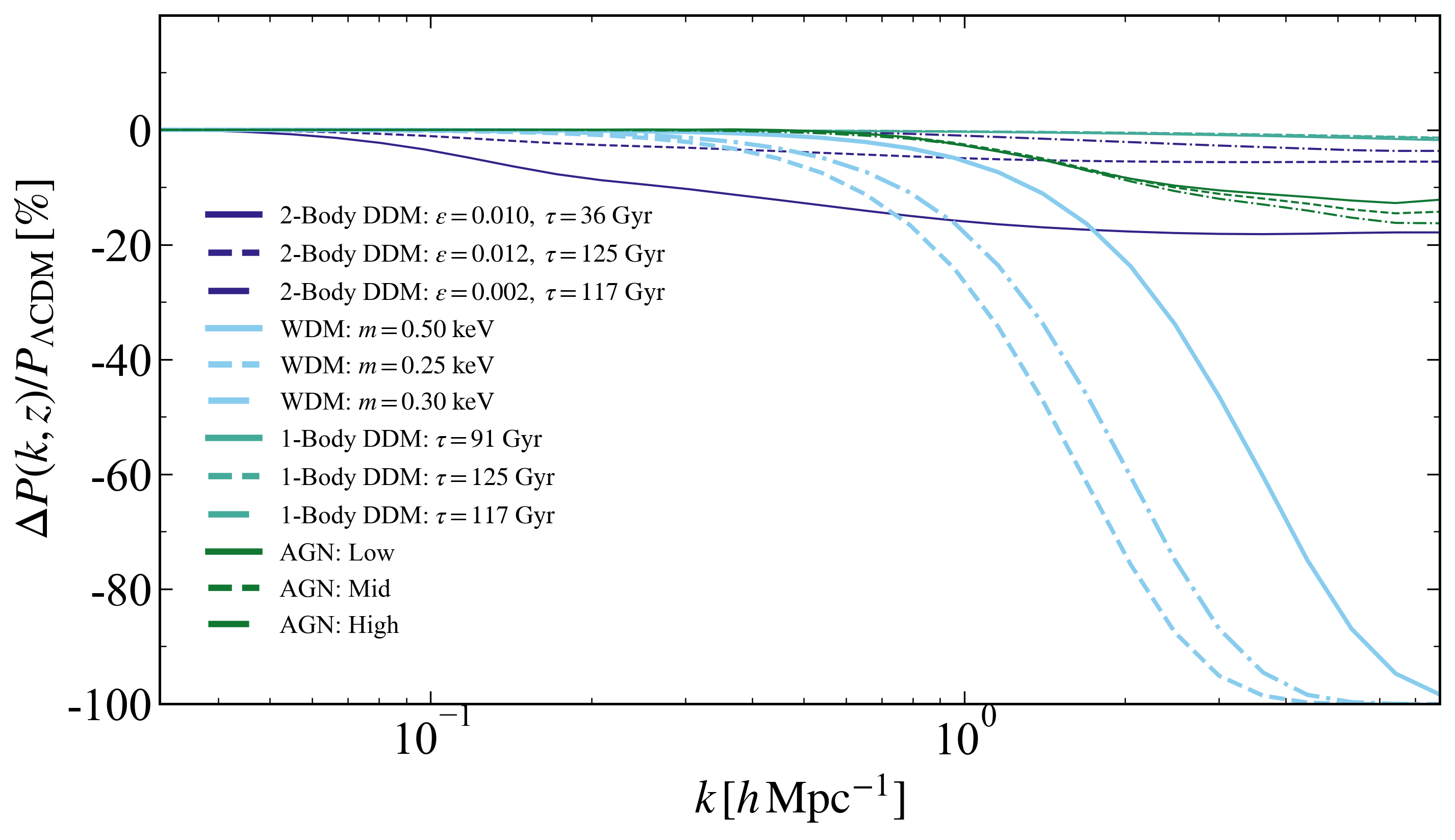}
 \caption {The small-scale suppression of power for DDM and other models at three redshifts $z =1.75, 2.00, 2.35$(top to bottom). The suppression of power for DDM may be indistinguishable from the similar effect arising from different models over a range of scales.}
 \label{fig:ps-supp}
\end{figure}

Fig. \ref{fig:ps-supp} shows the fractional difference of the power spectrum from alternative dark matter scenarios from the $\Lambda$CDM model.
At all these redshifts, we find that power suppression in the 2-body decaying DDM models could be confused with one-body decaying dark matter or warm dark matter scenarios over a wide range of scales. 

Other than such degeneracies with alternative dark matter scenarios, we also note that baryonic (AGN) feedback also leads to suppression of the matter power spectrum on these non-linear scales. 
Fig. \ref{fig:ps-supp} also shows the imprint of the AGN feedback using the  BCemu baryonic--correction emulator \cite{AGN_EMUL}. BCemu returns a redshift  and scale-dependent boost
$R_{\rm bary}(k,z)\!\equiv\!P_{\rm DM+bary}(k,z)/P_{\rm DM}(k,z)$ calibrated on a seven--parameter baryonification model and trained against hydrodynamical simulations). Within its calibrated range ($z\!\lesssim\!2$, $k\!\lesssim\!10\,h\,{\rm Mpc}^{-1}$), the emulator reproduces the simulation response at \mbox{$\sim$percent} accuracy (sub--percent for $z\!\lesssim\!1.5$; $\sim$2--3\% at $z\!=\!2$) and captures the characteristic $\mathcal{O}(10\%)$ suppression that emerges beyond $k\!\sim\!0.3\,h\,{\rm Mpc}^{-1}$ \cite{AGN_EMUL}.  
Here, we note that the scale dependence of the AGN response differs from our 2-body DDM models.  AGN suppression turns on gradually and steepens only for $k\gtrsim 0.3$--$0.5\,h\,{\rm Mpc}^{-1}$, whereas DDM imprints a smoother, earlier--onset reduction of power already in the quasi--linear window. The overlays in Fig.~\ref{fig:ps-supp} make this distinction explicit and show that plausible AGN feedback does not mimic the DDM signal on the large/intermediate modes that dominate our constraints, providing a clear path to marginalize over $R_{\rm bary}(k,z)$ in future inference \cite{AGN_EMUL}.

\begin{figure*}[ht!]
    \centering
    \includegraphics[width=0.32\textwidth, height=0.25\textwidth]{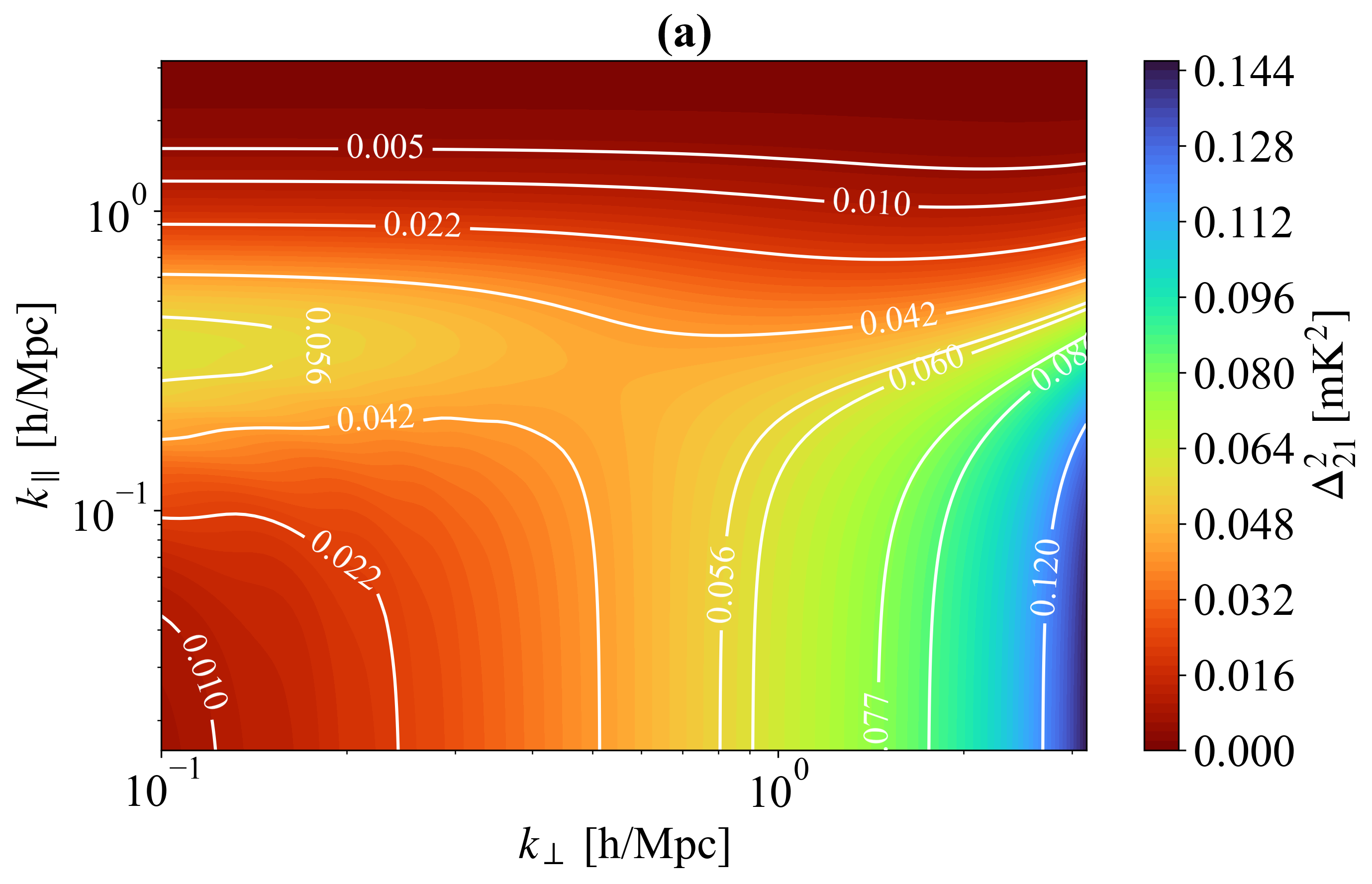}
    \includegraphics[width=0.32\textwidth, height=0.25\textwidth]{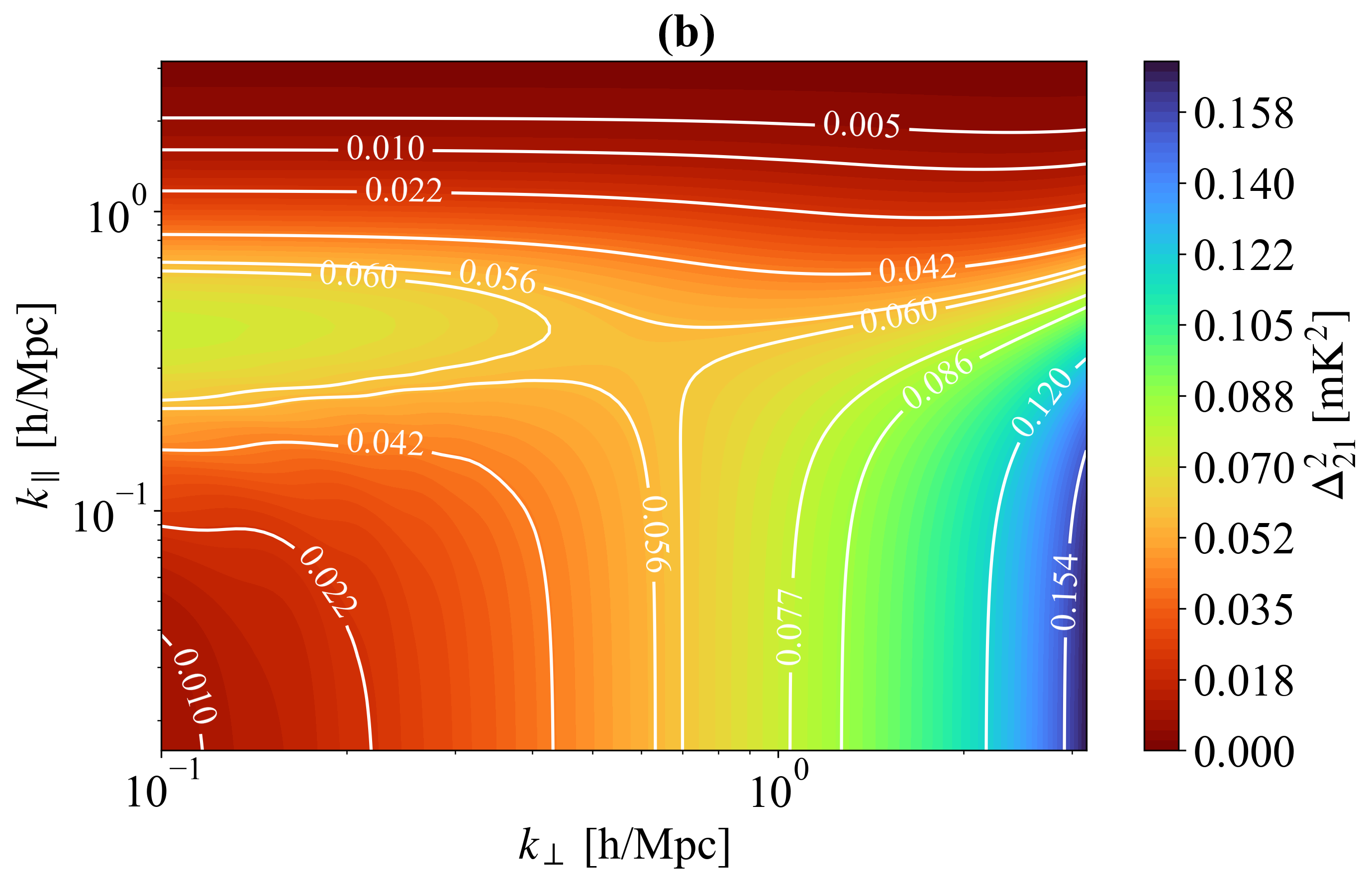}
    \includegraphics[width=0.32\textwidth, height=0.25\textwidth]{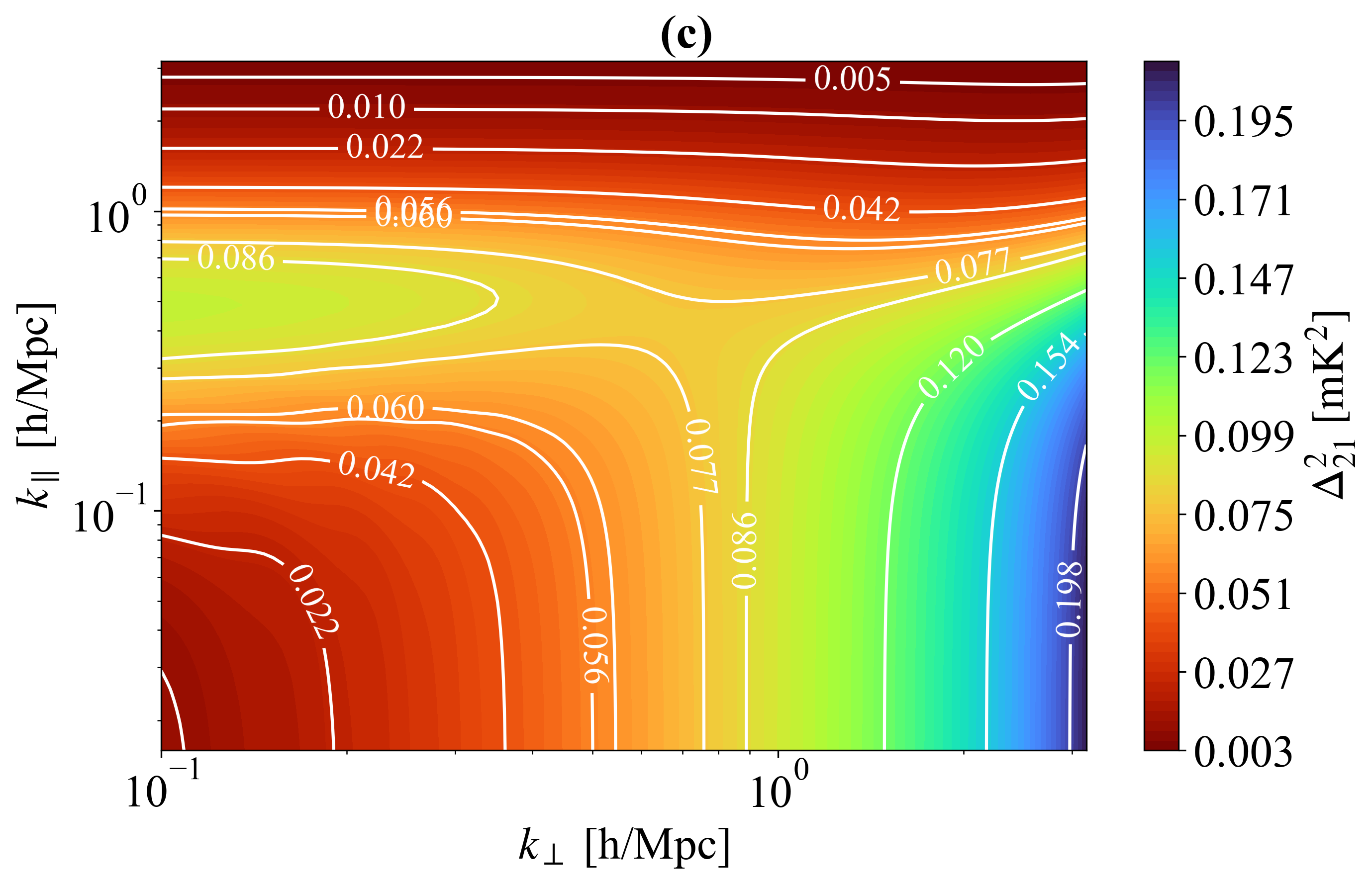}
    
\caption{(a),(b) and (c) shows the dimensionless anisotropic 21-cm power spectrum $k^3 P_{\rm HI} /2\pi^2$ for a DDM cosmology at the fiducial redshifts $z=1.75, 2.0$, and $2.35$ respectively at the fiducial values of parameters $(\Omega_m, \epsilon, \Gamma, H_0, f) = (0.3094, ~0.012,  ~0.008 {\rm Gyr^{-1}}, ~67.73 {\rm Km/s/Mpc}, ~1.0)$ respectively. The departure from spherical symmetry is seen on small scales. }
\label{fig:signal}
\end{figure*}

 The anisotropic 21-cm power spectrum $P_{\rm HI}$ is obtained by numerically evaluating the relevant cosmological quantities.
Fig.\ref{fig:signal}(a), (b), and (c) show the dimensionless anisotropic 21-cm power spectrum $k^3 P_{\rm HI} / 2\pi^2$ at the fiducial redshifts $z=1.75,~ 2.0,~ 2.35$, respectively, for a DDM cosmology with the given fiducial parameter values. We find that owing to the RSD and AP effect, the power spectrum shows a significant departure from spherical symmetry on small scales $ \approx k  \geq 1 {\rm Mpc}^{-1}$.

%\begin{figure*}[ht!]
%    \centering
%    \includegraphics[width=0.485\textwidth]{21cmDDM-1.pdf}
%     \includegraphics[width=0.485\textwidth]{SNR.pdf}
%    \caption{Figure (a) shows the  dimensionless anisotropic 21-cm power spectrum $k^3 P_{\rm HI} /2\pi^2$ for a DDM cosmology at the fiducial redshift $z=2$. The departure from spherical symmetry is seen on small scales. Figure (b) shows the signal to noise ratio $ P_{\rm HI}(k_{\parallel}, k_{\perp}, z) /   \delta  P_{\rm HI}(k_{\parallel}, k_{\perp}, z) $.}
%    \label{fig:signalandsnr}
%\end{figure*}

\section{Observational Considerations}

The noise estimates for the 21-cm auto power spectrum is obtained by considering a radio-interferometric observation.
We adopt the visibility-visibility correlation formalism \cite{mcquinn2006cosmological}  and assume that the visibility data is suitably binned  such that the correlation between baselines owing to the primary beam of the telescope can be ignored and the covariance matrix is diagonal.
Thus we  have for redshift $z$ corresponding to an observational frequency $\nu = 1420/(1+z)$MHz or wavelength $\lambda = 0.21(1+z)$m
\begin{equation}
\delta P_{\rm HI}(k, \mu, z)=\frac{P_{\rm HI}(k,\mu,z) +N_T(k, \mu , z)}{\sqrt{N_c}}
\end{equation}
where the first term corresponds to the cosmic variance, and the instrumental noise is  given by
\begin{equation}
N_T=\frac{\lambda^2 T_{s y s}^2 r^2 d r / d \nu}{A_e t_{\mathbf{k}}}
\end{equation}
Here $r$ denotes the comoving distance to the source, $A_e$ is the effective area of the individual antenna dish, $T_{sys}$ is the system temperature assumed to be dominated by the sky temperature, and 
\begin{equation}
t_{\mathbf{k}}=T_o N_{a n t}\left(N_{a n t}-1\right) A_e \rho / 2 \lambda^2  \end{equation}
is the fraction of the total observation time $T_o$ for a given pointing spent on the given mode. We have considered a radio-array with $N_{ant}$ antennae spread out in a plane, such that the total number of visibility pairs $N_{ant} ( N_{ant} -1) /2$ are distributed over different baselines according to a normalized baseline distribution function $\rho(k_{\perp}, \nu)$.
The noise is suppressed by a factor $\sqrt {N_c}$ where $N_c$ is the number of modes in a given survey volume. Thus we have 
\begin{equation}
N_c=2 \pi k^2 \Delta k \Delta \mu r^2(dr / d \nu) B \lambda^2 / A_e(2 \pi)^3    
\end{equation}
For $N_{point}$ different pointings, the noise is further suppressed by a factor $1/\sqrt{N_{point}}$.
\begin{figure*}[ht!]
    \centering
    \includegraphics[width=0.6\textwidth]{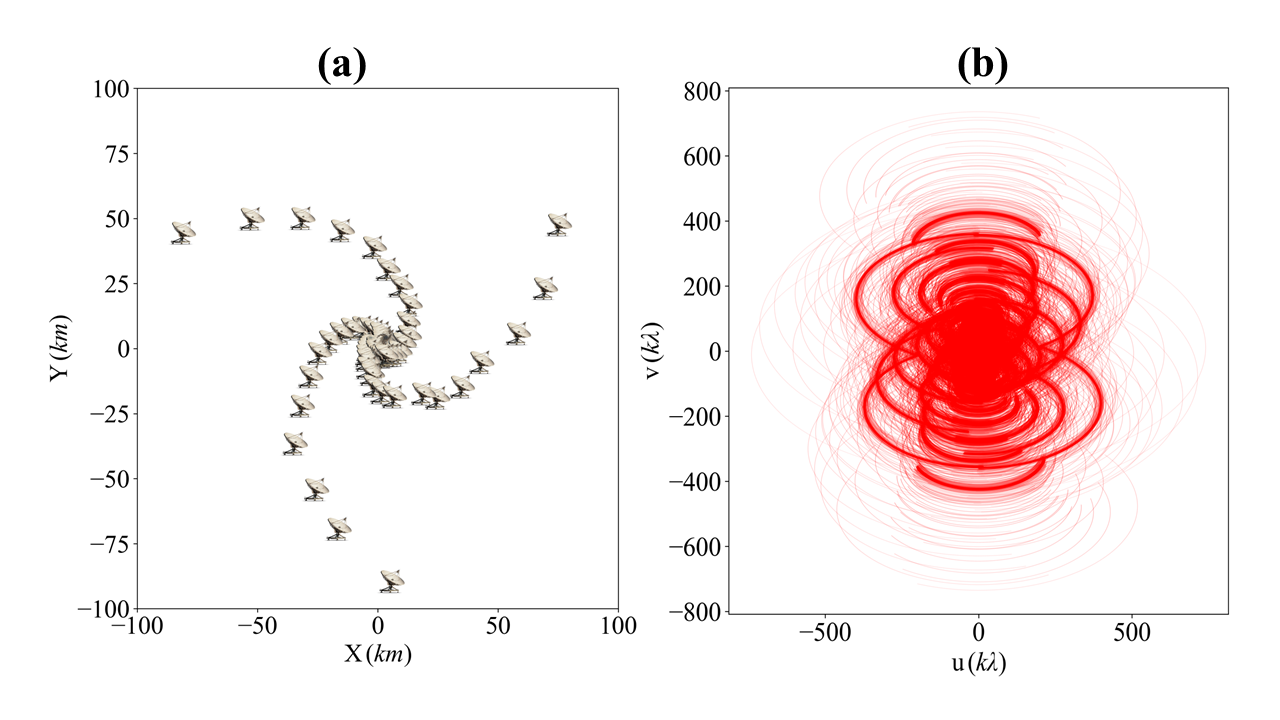}
       \caption{(a) Shows the actual antenna distribution. (b) Shows the sampling $u-v$ coverage for $5$hrs synthesis imaging.}
    \label{fig:arraylayout}
\end{figure*}

\begin{figure*}[ht!]
    \centering
        \includegraphics[width=0.253\textwidth]{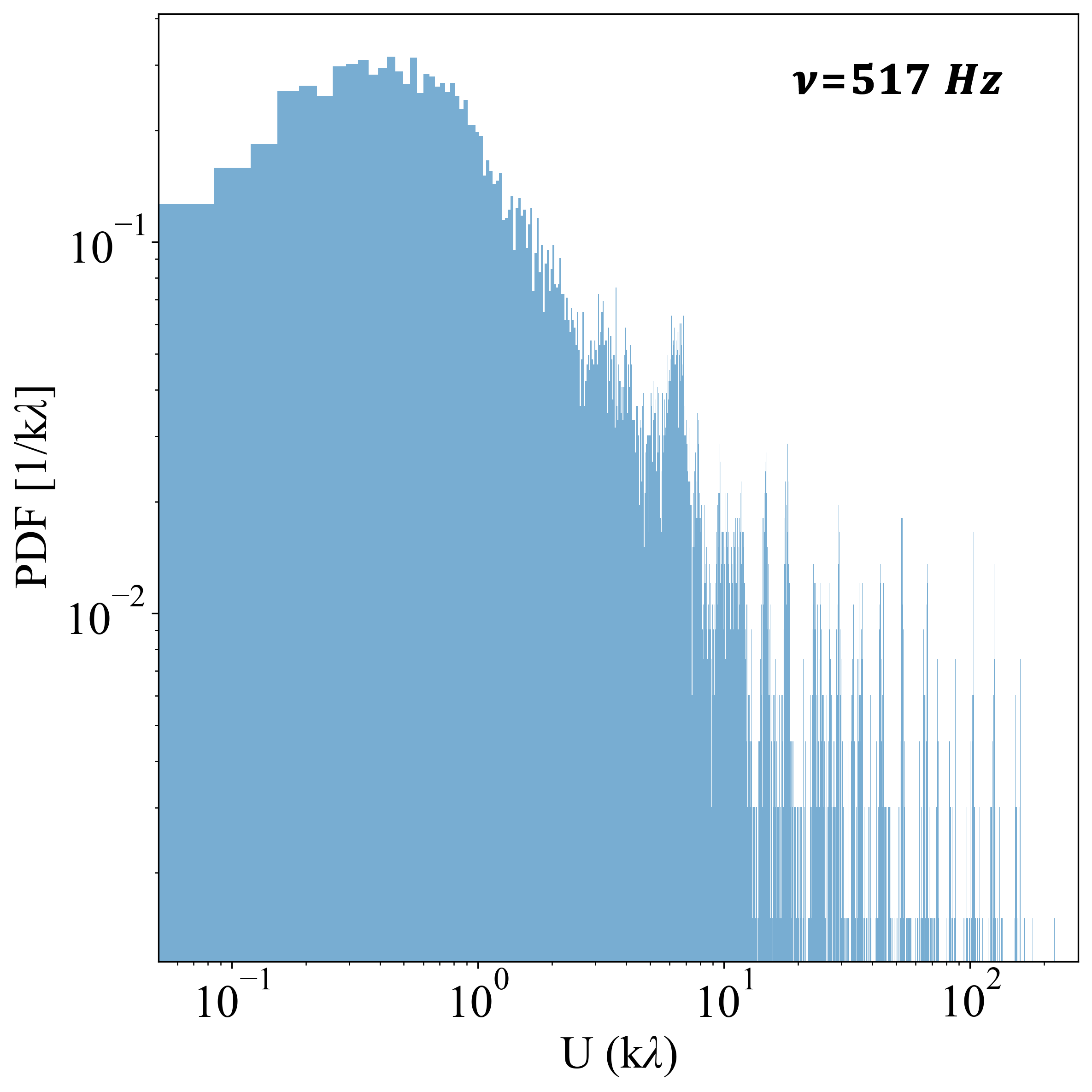}
        \includegraphics[width=0.253\textwidth]{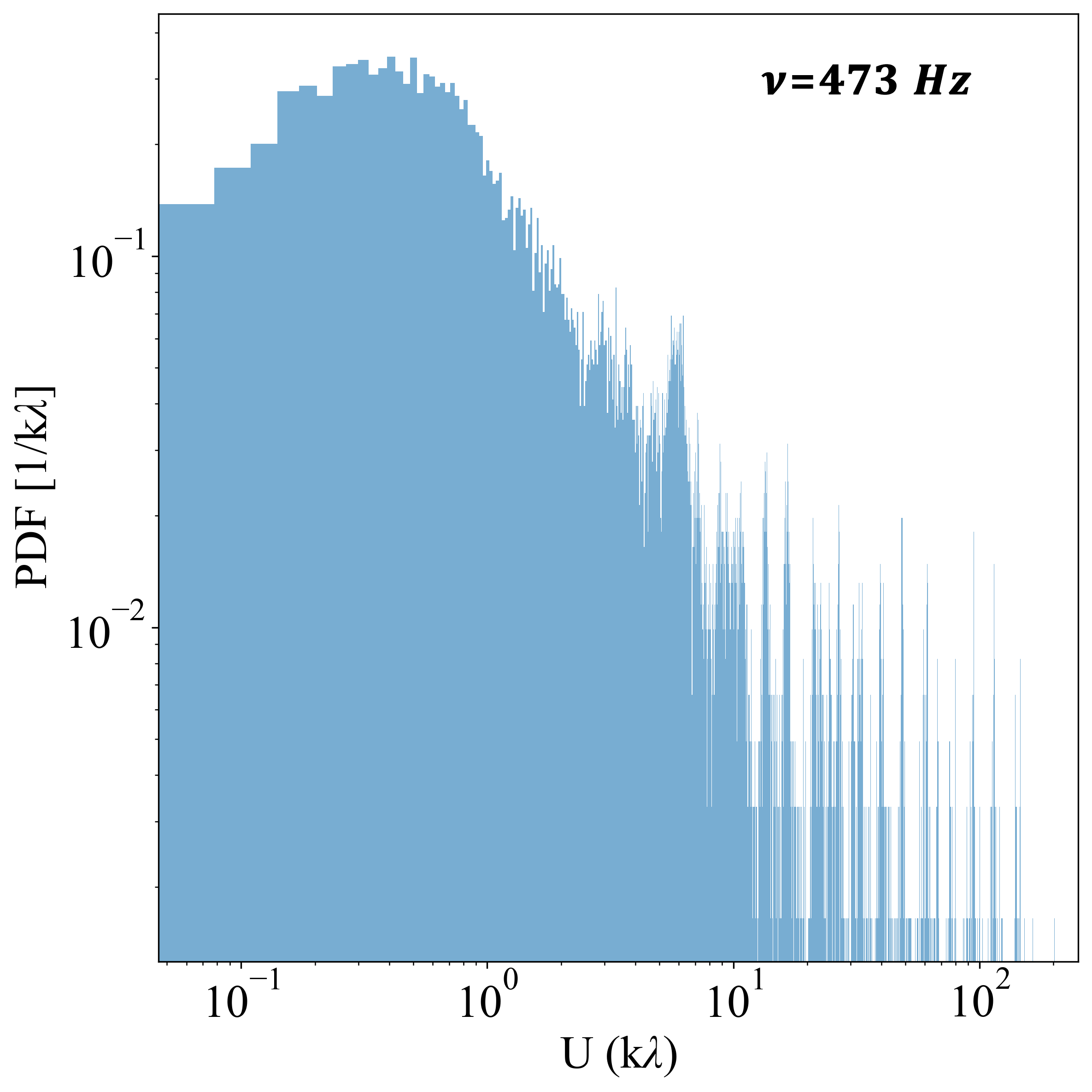}
        \includegraphics[width=0.253\textwidth]{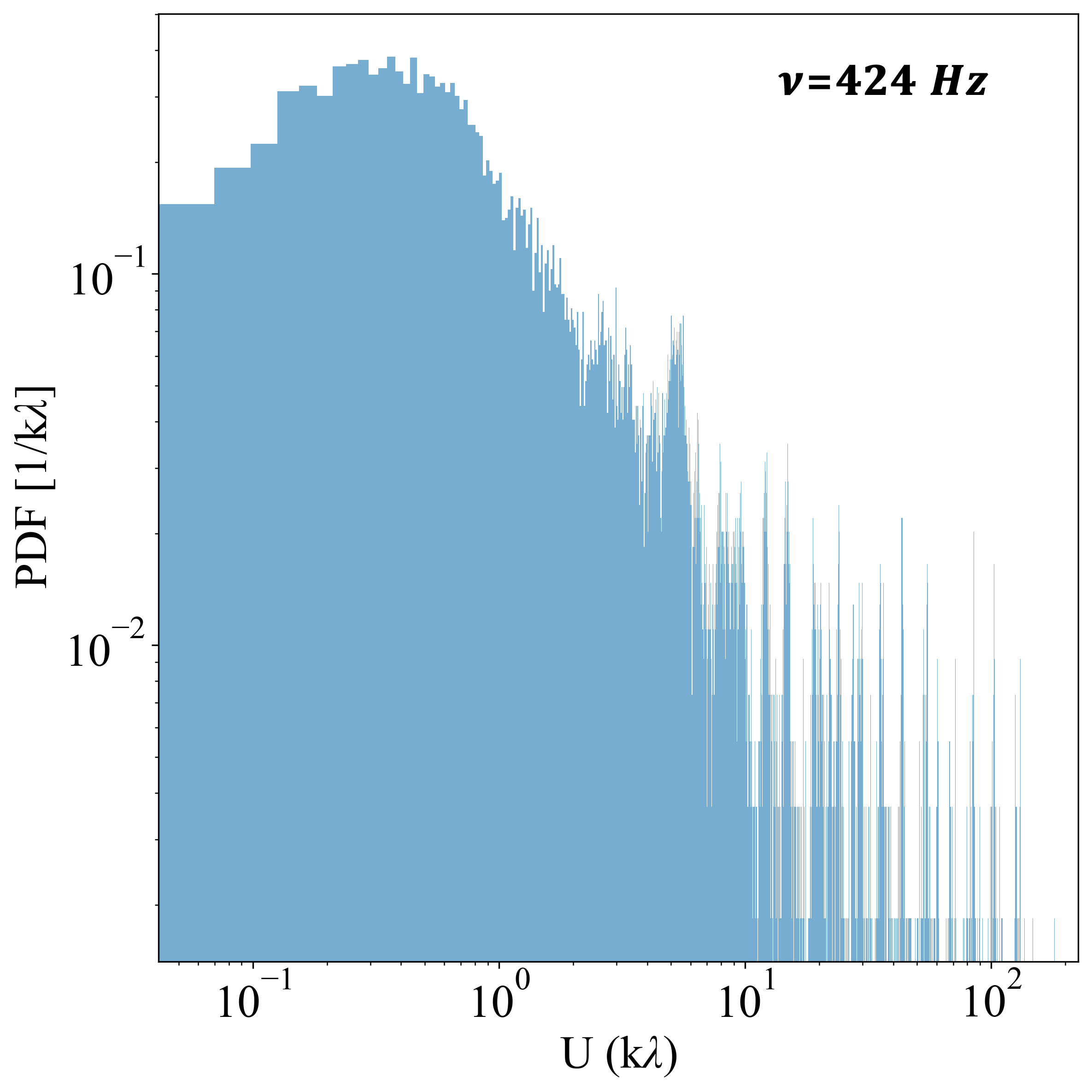}
    \caption{The baseline distribution function $\rho(U)$ at the three observing frequencies. }
    \label{fig:pdf}
\end{figure*}

The noise estimates are based on a futuristic SKA1-Mid like intensity mapping experiment. We consider an interferometer with $197$ dish antennae, each of diameter $15$m
and antenna efficiency of $0.7$. The array specifications are obtained from \href{SKA webpage}{https://www.skao.int/en}. The baseline distribution function is obtained by assuming that the antennae are  distributed with a density which decreases with radial distance from the center. Fig. \ref{fig:arraylayout} (a) shows the actual antenna positions. The corresponding $(u-v)$ coverage is shown in  Fig.\ref{fig:arraylayout} (b) for synthesis interferometry.
The normalized baseline distribution function $\rho$ is shown in Fig.\ref{fig:pdf}. We find that while large scales (small $(u,v)$) are densely sampled, the sampling is poor for small-scale modes. This increases the noise on small scales due to poor visibility sampling. We consider about
a $128$MHz bandwidth observation at three observing frequencies of $517$MHz, $473$MHz, and $424$MHz corresponding to the probing redshifts of 
$z=1.75$, $z=2.0$ and $z=2.35$ respectively. The field of view at these redshifts are $2.21^{\circ}$, $2.407^{\circ}$, $2.68^{\circ}$ respectively. We also consider a $k-$ binning with $\Delta k = k/N_{bin} \ln(k_{max}/k_{min})$, with $N_{bin} = 40$ and where $k_{max}$ and $k_{min}$ are set by the largest baseline and the dish size, respectively. We
consider a total of  $T=4000$ hrs observations for $10$ pointings with $400$ hrs per pointing.

\begin{table}[h!]
\centering
\caption{Telescope parameters used in our analysis.}
\label{tab:telescope-params}

% ──────────────────────────────────────────────────────────────────────────────
% First subtable: array-level parameters
% ──────────────────────────────────────────────────────────────────────────────
\begin{tabular}{cccccc}
\hline
$N_{\mathrm{ant}}$ & Antennae efficiency & $D_{\mathrm{dish}}$ 
  & $T = T_{0}\,N_{\mathrm{point}}$ & $T_{\mathrm{sys}}$ & $B$ \\
\hline
197 & 0.7 & 15\,m & 4000\,hrs & 60\,K & 128\,MHz \\
\hline
\end{tabular}

% ──────────────────────────────────────────────────────────────────────────────
% Second subtable: frequency-dependent parameters
% ──────────────────────────────────────────────────────────────────────────────
\begin{tabular}{cccc}
\hline
Frequency (MHz) & FoV 
  & $k_{\min}\,\bigl(\mathrm{h\,Mpc}^{-1}\bigr)$ 
  & $k_{\max}\,\bigl(\mathrm{h\,Mpc}^{-1}\bigr)$ \\
\hline
517 & $2.21^{\circ}$  & 0.0955  & 7.238  \\
473 & $2.407^{\circ}$ & 0.0811  & 6.634  \\
424 & $2.68^{\circ}$  & 0.0667  & 5.941  \\
\hline
\end{tabular}
\end{table}

%\begin{figure}[ht!]
%    \centering
%    \includegraphics[width=0.65\textwidth]{cornerplot.pdf}
%    \caption{Error projections at the $68 \%$ and $95 \%$ and $99\%$ confidence levels for the %parameters ($H_o$, $\Gamma$, $\epsilon$, $f$)}.
%    \label{fig:contour}
%\end{figure}

To account for the discrete nature of the sources, one must add the Poisson shot noise contribution 
\begin{equation} 
P_{SN}(k,z) = \frac{\mathcal{C}^2(z)}{b_T^2} \frac{\int_0^{\infty} n(M,z) M_{HI}^2 (M, z) dM }{\int_0^{\infty} n(M,z) M_{HI} (M, z) dM }
\end{equation}
where $n(M,z)$ is the Halo mass function and $M_{HI}$ is the HI mass in a halo of mass $M$, given by a fit function \cite{shotnoise}
\begin{equation}
M_{HI} = M_0 \left ( \frac{M}{M_{min}} \right ) ^\alpha
e^{ -(M_{min}/ M )^{0.35}} 
\end{equation}
where the fit parameters $(M_{min}, M_0, \alpha) $ are adopted at the given redshifts from \cite{shotnoise}.

\section{Results and Discussion}
\begin{figure*}[ht!]
    \centering
    \includegraphics[width=0.32\textwidth, height=0.25\textwidth]{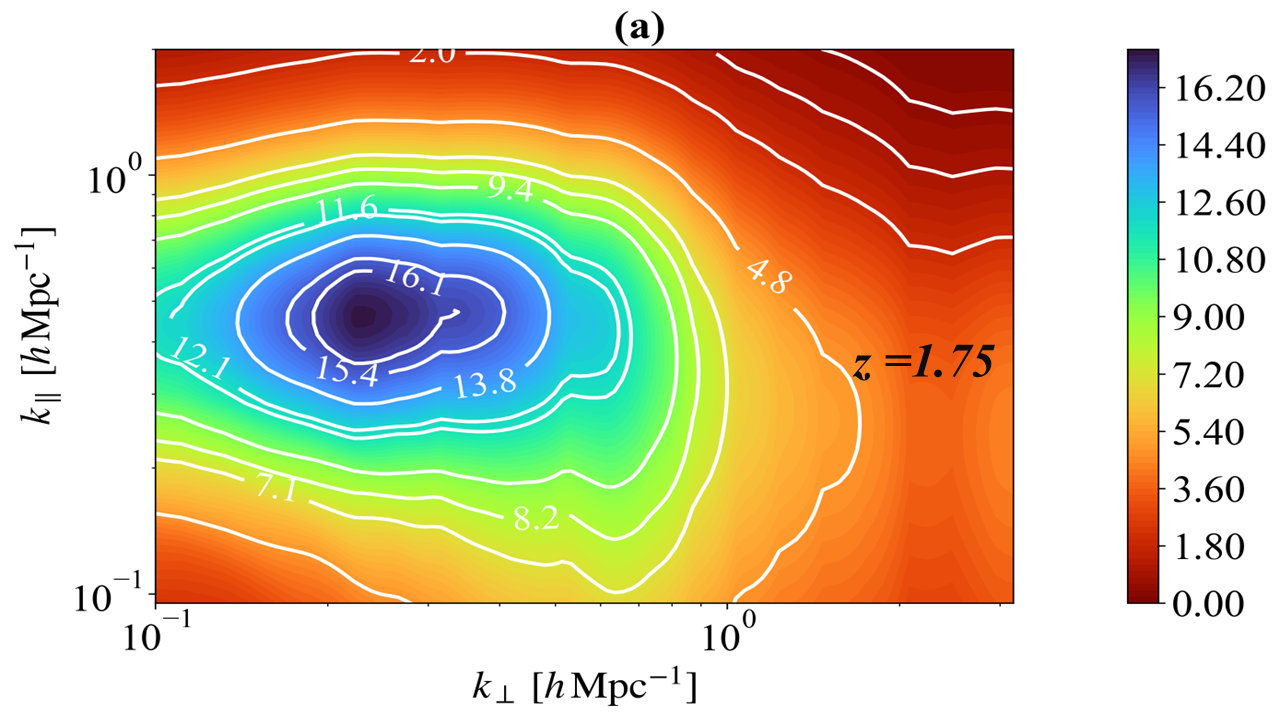}
    \includegraphics[width=0.32\textwidth, height=0.25\textwidth]{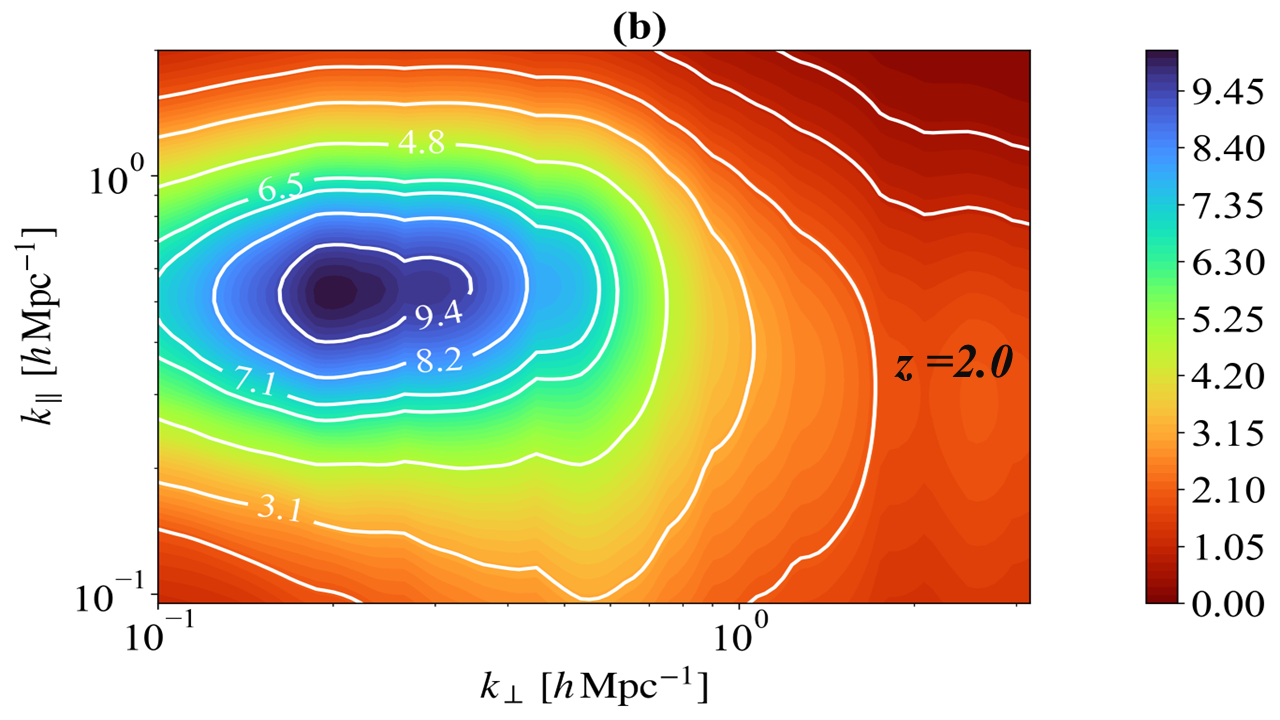}
    \includegraphics[width=0.32\textwidth, height=0.25\textwidth]{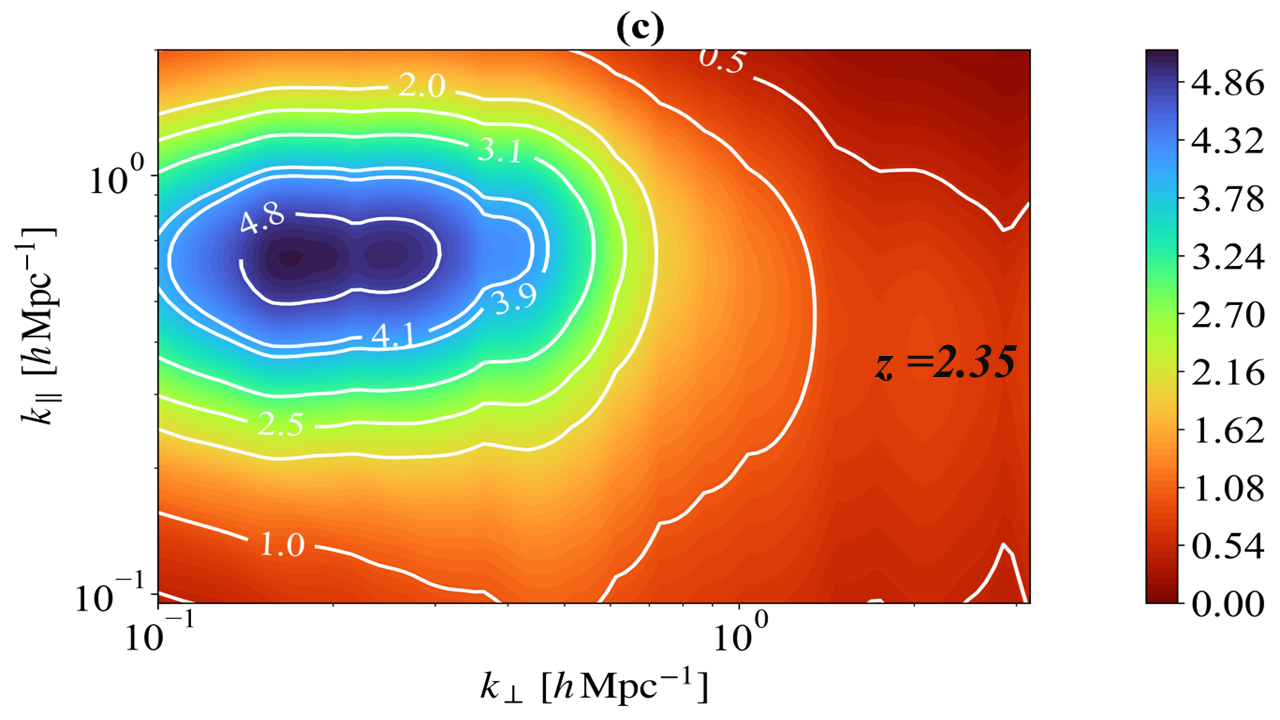}
    \includegraphics[width=0.32\textwidth, height=0.25\textwidth]{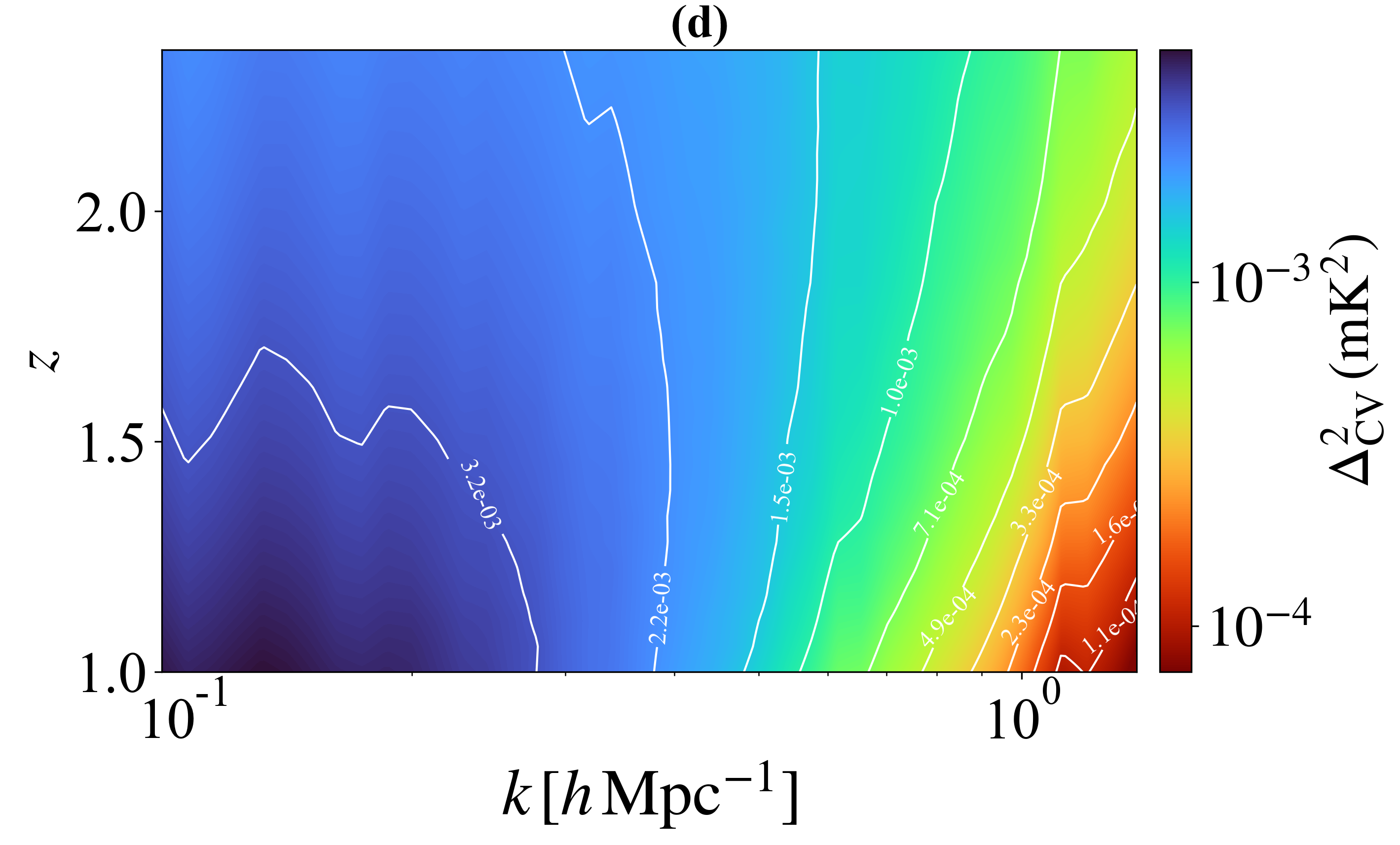}
    \includegraphics[width=0.32\textwidth, height=0.25\textwidth]{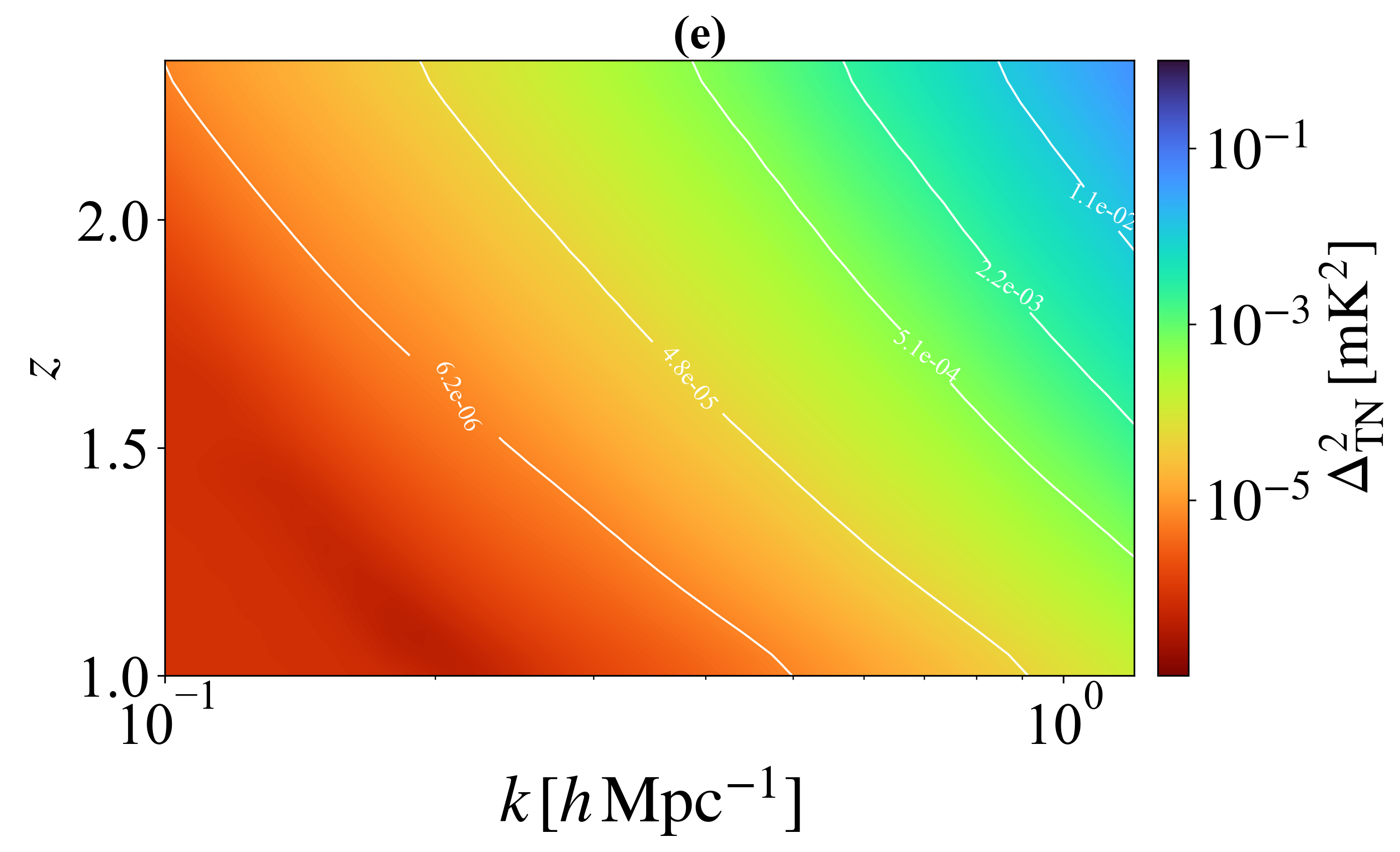}
    \includegraphics[width=0.32\textwidth, height=0.26\textwidth]{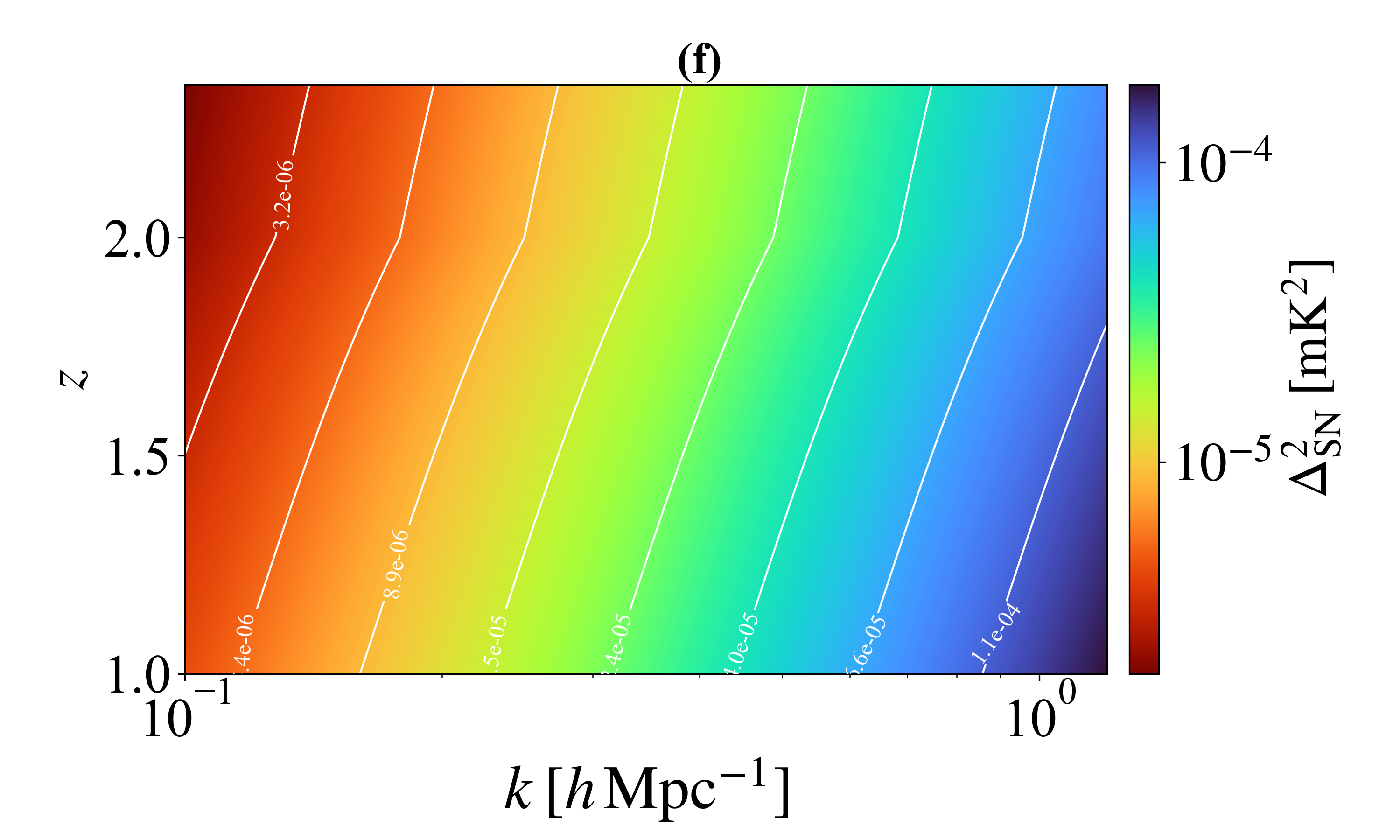}
    \caption {Figures (a),(b) and (c) shows the signal to noise ratio $ P_{\rm HI}(k_{\parallel}, k_{\perp}, z) /   \delta  P_{\rm HI}(k_{\parallel}, k_{\perp}, z) $ at the fiducial redshifts $z=1.75, 2.0, 2.35$ respectively. The figures (d),(e),(f) shows the behaviour of Cosmic variance noise, Thermal Instrumental Noise and Shot Noise due to discrete sampling respectively in the $k-z$ plane and  demonstrates which noise component dominates at which scales and redshifts.}
    \label{fig:snr}
\end{figure*}

    \begin{figure*}[ht!]
    \centering
    \includegraphics[width=0.3\textwidth,  height=0.25\textwidth ]{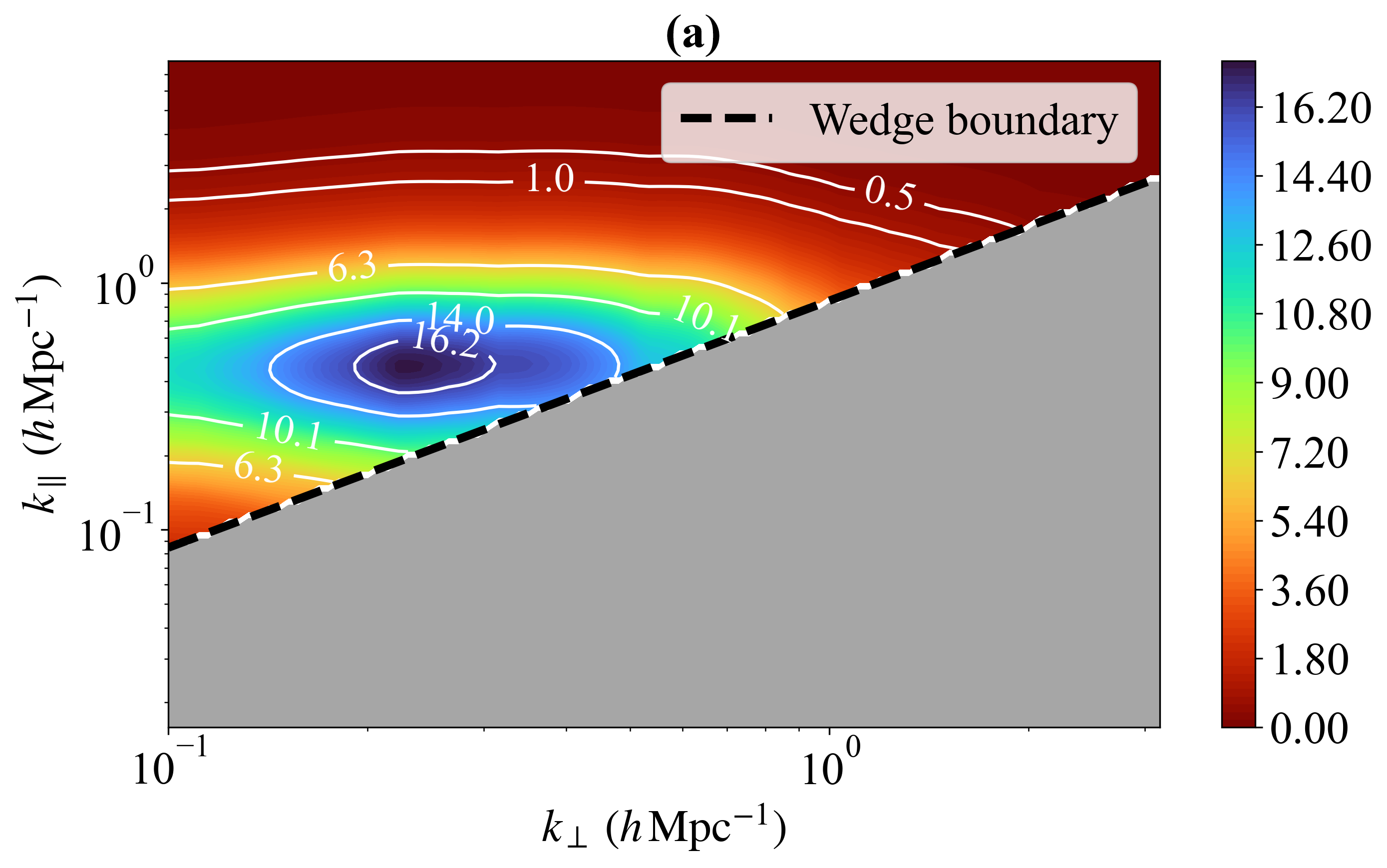}
    \includegraphics[width=0.3\textwidth,  height=0.25\textwidth]{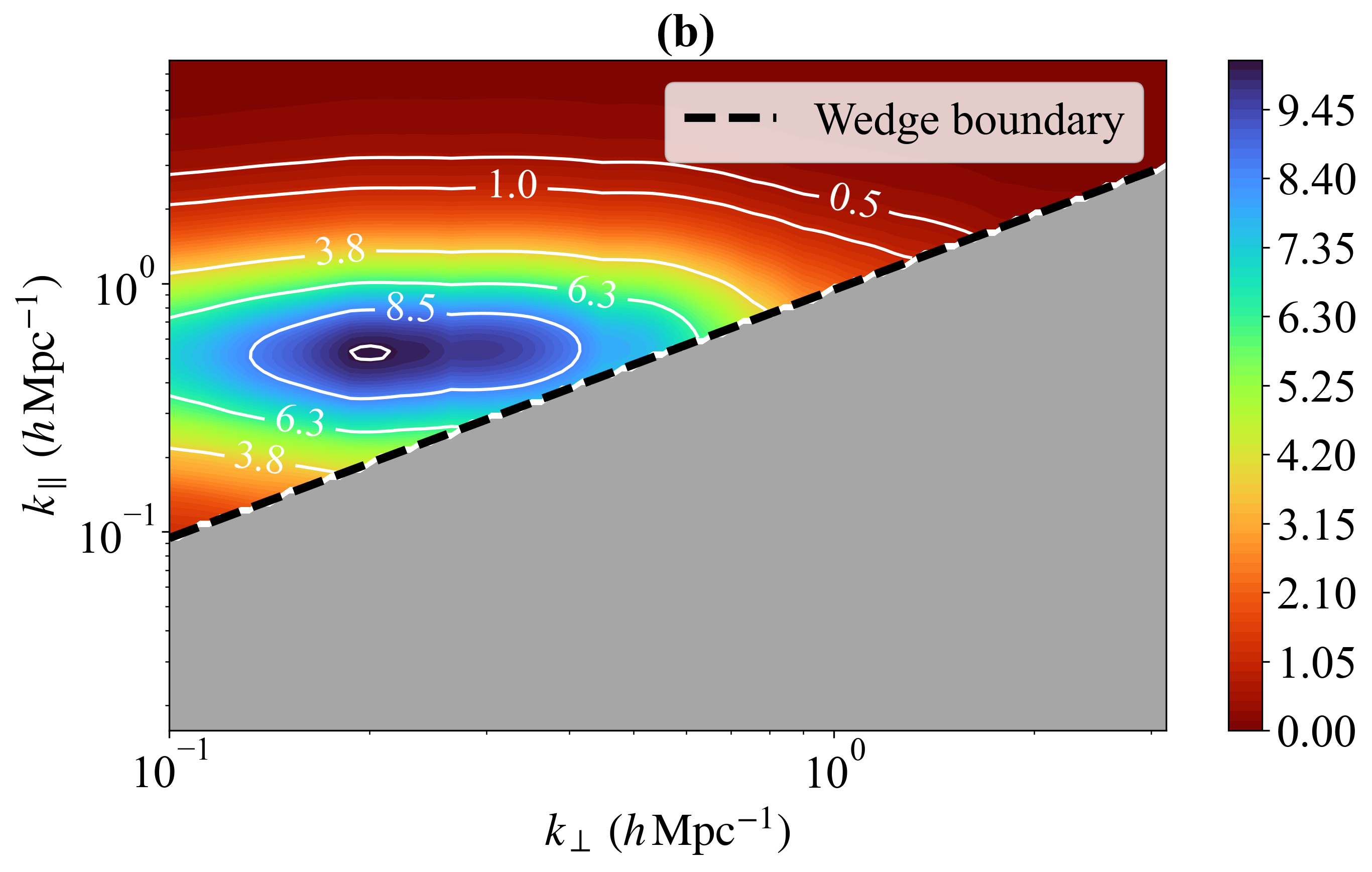}
    \includegraphics[width=0.3\textwidth,  height=0.25\textwidth]{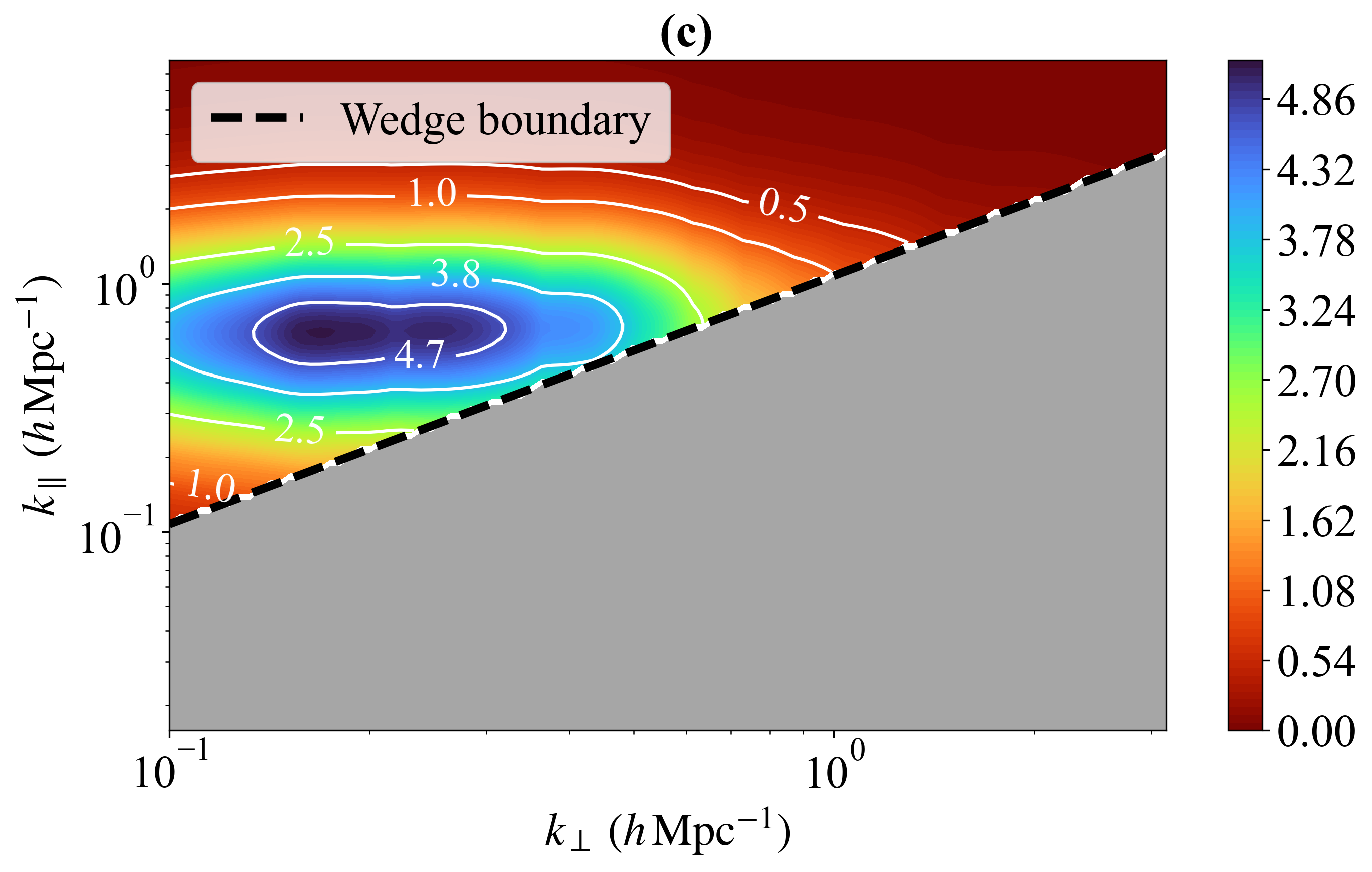}
    \caption{Foreground wedge is shown in the SNR plot at the three observing redshifts (a) $ z = 1.75$, (b) $z= 2.0$ (c) $z= 2.35$}
    \label{fig:wedge}
\end{figure*}

 Fig.\ref{fig:snr} (a),(b) and (c) shows the signal to noise ratio for the anisotropic 21-cm power spectrum in the 
$(k_{\parallel}, k_{\perp}) $ space for the three redshifts under consideration. 
Considering $SNR> 5$ to be a  threshold for detection, we find that for $z=1.75$
the $k$-space region where this occurs is 
approximately given by 
$0.052 {\rm Mpc}^{-1}\leq k_{\parallel} \leq 1.28 {\rm Mpc}^{-1}$,   
$0.1 {\rm Mpc}^{-1}\leq k_{\perp} \leq 1.52 {\rm Mpc}^{-1}$.
For $z = 2$, this region is 
$0.221 {\rm Mpc}^{-1}\leq k_{\parallel} \leq 1.134 {\rm Mpc}^{-1}$,   
$0.1 {\rm Mpc}^{-1}\leq k_{\perp} \leq 0.73 {\rm Mpc}^{-1}$ and for $z=2.35$,this  region is 
$0.53 {\rm Mpc}^{-1}\leq k_{\parallel} \leq 0.734 {\rm Mpc}^{-1}$,   
$0.152 {\rm Mpc}^{-1}\leq k_{\perp} \leq 0.3 {\rm Mpc}^{-1}$.
Outside this range the SNR is $<5$ and would not entail a statistically significant detection. While poor visibility sampling is the cause of the cosmic variance noise on large scales, it is the instrument noise that dominates on small scales.

Fig. \ref{fig:snr} (d),(e) and (f) shows the individual noise contributions to the power spectrum in $(k, z)$ plane corresponding to Cosmic variance, Thermal Noise and Shot noise respectively.  For this figure, we have considered the spherically averaged power spectrum.  
Fig. \ref{fig:snr} (d) which shows the cosmic variance noise given by $k^3/(2 \pi^2) P_{HI} / \sqrt{N_c}$ is very large on large scales and degraded SNR at small $k$ due to lack of modes. The cosmic variance is subdominant on small scales.
The scale dependence of the Thermal noise owes its origin to the baseline distribution and dominates at small scales where the baseline coverage is poor. Thermal noise is high at large redshifts. 
The shot noise contribution (multiplied by  $k^3/2\pi^2\sqrt N_c$ for comparison) shown in Fig. \ref{fig:snr}(f)  is calculated using the formalism using the halo-model prescription of \cite{shotnoise} and is evaluated using the best-fit $M_{\rm HI}(M,z)$ relation with interpolated redshift-dependent parameters from their Table~1. For $n(M,z)$, we use the Sheth--Tormen mass function. 
The shot noise is very subdominant, especially at higher redshifts, and this allows for the extraction of significant cosmological information even on small scales.

Having established the possibility of detecting the signal, we now make error estimates on the DDM  parameters  $q_m = (\epsilon, \Gamma, H_o, f)$ for the proposed 21-cm intensity mapping experiment described above. 
The sensitivity of the 21-cm power spectrum given in Eq.\ref{eq:21cmps} to DDM parameters occurs due to the suppression of power on small scales and also due to the change in the background expansion rate $H(z)$ that affects the normalization $\mathcal{C} (z)$ and the anisotropy of the power spectrum.
The dependence of the parameters $\{ q_m \}$  on $H(z)$ given in Eq. \ref{eq:hubble}
is through the equations for $\rho_i(z) $ ( Eq: 2.2-2.5 ), which we integrate numerically. The power spectrum $P(k,z)$  is obtained using the emulator \cite{Giri(2024)}.
The 21-cm power spectrum $P_{HI}(k,z,\mu)$ is numerically computed for a set of parameter values $\{ q_m \}$. 
\begin{table*}[t]
  \centering
  \small
  \renewcommand{\arraystretch}{1.1}
  \setlength{\tabcolsep}{4pt}
  \begin{tabular}{l c c c c}
    \hline
    \hline 
    \textbf {TABLE  A} & && & \\
    \hline 
    \textbf{Parameters}
      & $\epsilon$
      & $\Gamma\,(\mathrm{Gyr}^{-1})$
      & $H_{0}\,(\mathrm{km/s/Mpc})$
      & $f$ \\
    \hline
    \textbf{1\,$\sigma$ error}
      & $0.012 \pm 4.051\times10^{-4}$
      & $0.008 \pm 5.509\times10^{-4}$
      & $67.73 \pm 0.1406$
      & $1.0 \pm 0.08$ \\
     \hline
      &&&&\\
    \hline
    \textbf {TABLE  B} & && & \\
  \hline
  \textbf{Parameters} 
    & $\epsilon$ 
    & $\Gamma$ (Gyr$^{-1}$) 
    & $H_{0}$ (km/s/Mpc) 
    & $f$ \\
  \hline
  \textbf{1$\sigma$ error} 
    & $0.012 \pm 1.19 \times 10^{-3}$ 
    & $0.008 \pm 9.63 \times 10^{-4}$ 
    & $67.73 \pm 0.29$ 
    & $1.0 \pm 0.10$ \\
   \hline
      &&&&\\
    \hline
    \textbf {TABLE  C} & && & \\
  \textbf{Parameters} 
    & $\epsilon$ 
    & $\Gamma$ (Gyr$^{-1}$) 
    & $H_{0}$ (km/s/Mpc) 
    & $f$ \\
  \hline
  \textbf{1$\sigma$ error} 
    & $0.012 \pm 1.69 \times 10^{-1}$ 
    & $0.008 \pm 2.97 \times 10^{-1}$ 
    & $67.73 \pm 6.17$ 
    & $1.0 \pm 1.11 \times 10^{1}$ \\
   \hline
      &&&&\\
    \hline
   \textbf {TABLE  D} & && & \\
    \textbf{Parameters}
      & $\epsilon$
      & $\Gamma\,(\mathrm{Gyr}^{-1})$
      & $H_{0}\,(\mathrm{km/s/Mpc})$
      & $f$ \\
    \hline
    \textbf{1\,$\sigma$ error}
      & $1.14 \times 10^{-4} \pm 8.897\times10^{-6}$
      & $1.04 \times 10^{-3} \pm 8.680\times10^{-5}$
      & $67.73 \pm 0.142$
      & $1.0 \pm 0.01219$ \\
      \hline
      &&&&\\
    \hline
 
   \textbf {TABLE  E} & && & \\
    \hline
    \textbf{Parameters}
      & $\epsilon$
      & $\Gamma\,(\mathrm{Gyr}^{-1})$
      & $H_{0}\,(\mathrm{km/s/Mpc})$
      & $f$ \\
    \hline
    \textbf{1\,$\sigma$ error}
      & $1.14 \times 10^{-4} \pm 2.712\times10^{-5}$
      & $1.04 \times 10^{-3} \pm 1.489\times10^{-4}$
      & $67.73 \pm 0.175$
      & $1.0 \pm 0.27$ \\
    \hline
    \hline
  \end{tabular}
  \caption{A : Fiducial values and $1\sigma$ uncertainties for parameters $(\epsilon, \Gamma, H_0, f)=(0.012, 0.008 {\rm Gyr^{-1}},67.73 {\rm Km/s/Mpc}, ~1.0)$. Foregrounds are absent in this analysis. B : Fiducial values and 1$\sigma$ uncertainties for parameters $(\epsilon, \Gamma, H_0, f)$$=$$(0.012, 0.008 {\rm Gyr^{-1}},67.73 {\rm Km/s/Mpc}, ~1.0)$. The foregrounds are present in this analysis. ~C: Fiducial values and updated 1$\sigma$ uncertainties for the parameters $(\epsilon, \Gamma, H_0, f)$$=$$(0.012, 0.008 {\rm Gyr^{-1}},67.73 {\rm Km/s/Mpc}, ~1.0)$ when the bias is marginalized. ~D: Fiducial values and $1\sigma$ uncertainties for parameters $(\epsilon, \Gamma, H_0, f)$$=$$(1.14\times10^{-4}, 1.04 \times10^{-3} {\rm Gyr^{-1}},67.73 {\rm Km/s/Mpc}, ~1.0)$. Foregrounds are absent in this analysis. ~E: Fiducial values and $1\sigma$ uncertainties for parameters $(\epsilon, \Gamma, H_0, f)$$=$$(1.14\times10^{-4},1.04\times10^{-3}{\rm Gyr^{-1}},67.73 {\rm Km/s/Mpc},1.0)$. Foregrounds are present in this analysis.}
  \label{tab:1}
\end{table*}

The Fisher matrix  
\begin{equation}
F_{m n}=\sum_{k, \mu} \frac{1}{\delta P_{\rm HI} (k, \mu, z)^2} \frac{\partial P_{\rm HI} (k, \mu, z)}{\partial q_m} \frac{\partial P_{\rm HI} (k, \mu, z)}{\partial q_n} 
\end{equation}
allows us to put errors on the parameters from the Cramer-Rao bound  $\delta q_m = \sqrt{F_{m m}^{-1}}$, assuming a Gaussian likelihood. 
The matrix elements $F_{mn}$ are obtained numerically by taking derivatives of 
$P_{HI}(k,z,\mu)$ with respect to the parameters around the fiducial values.

\begin{figure*}[t]
    \centering
    \includegraphics[width=0.33\textwidth]{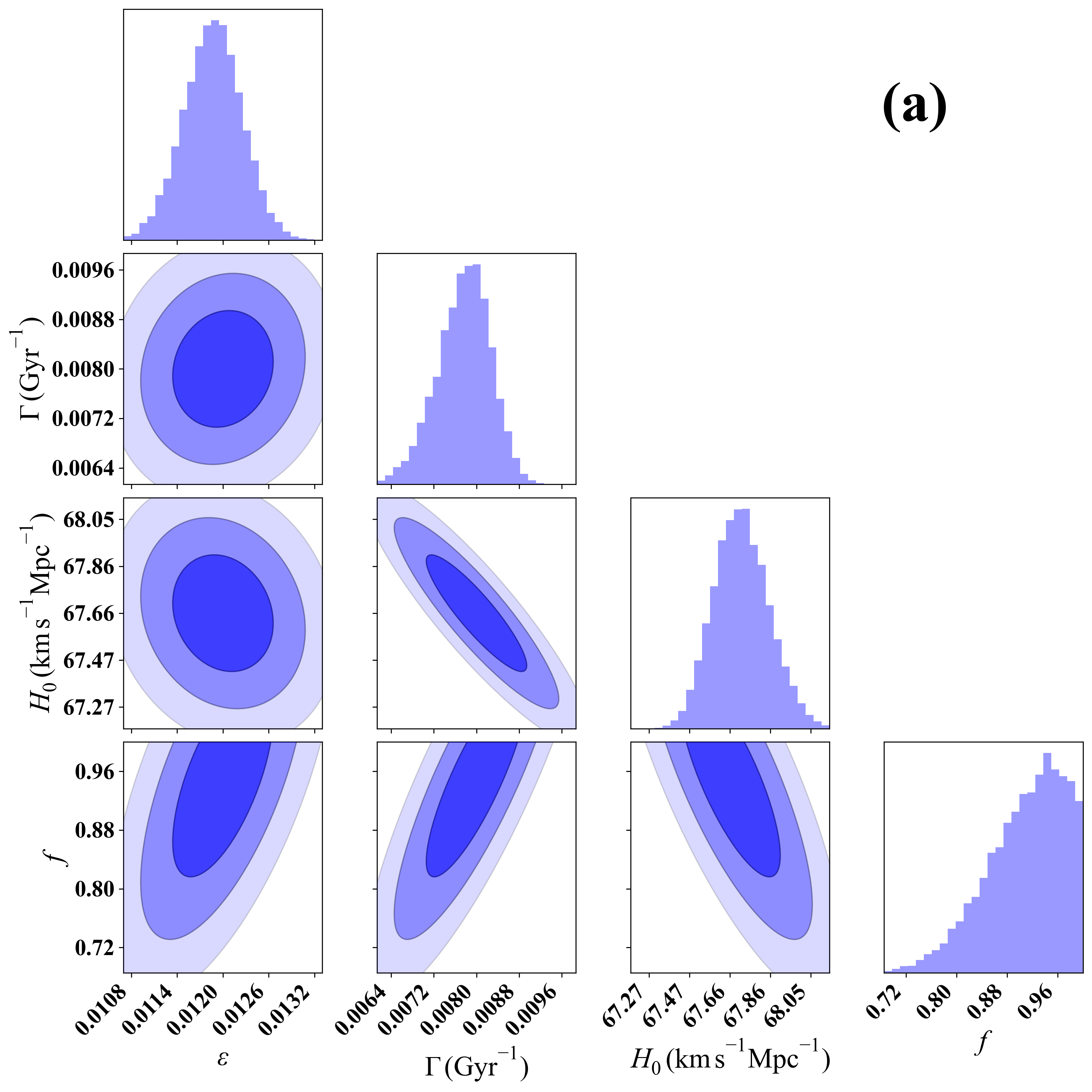}
    \includegraphics[width=0.33\textwidth]{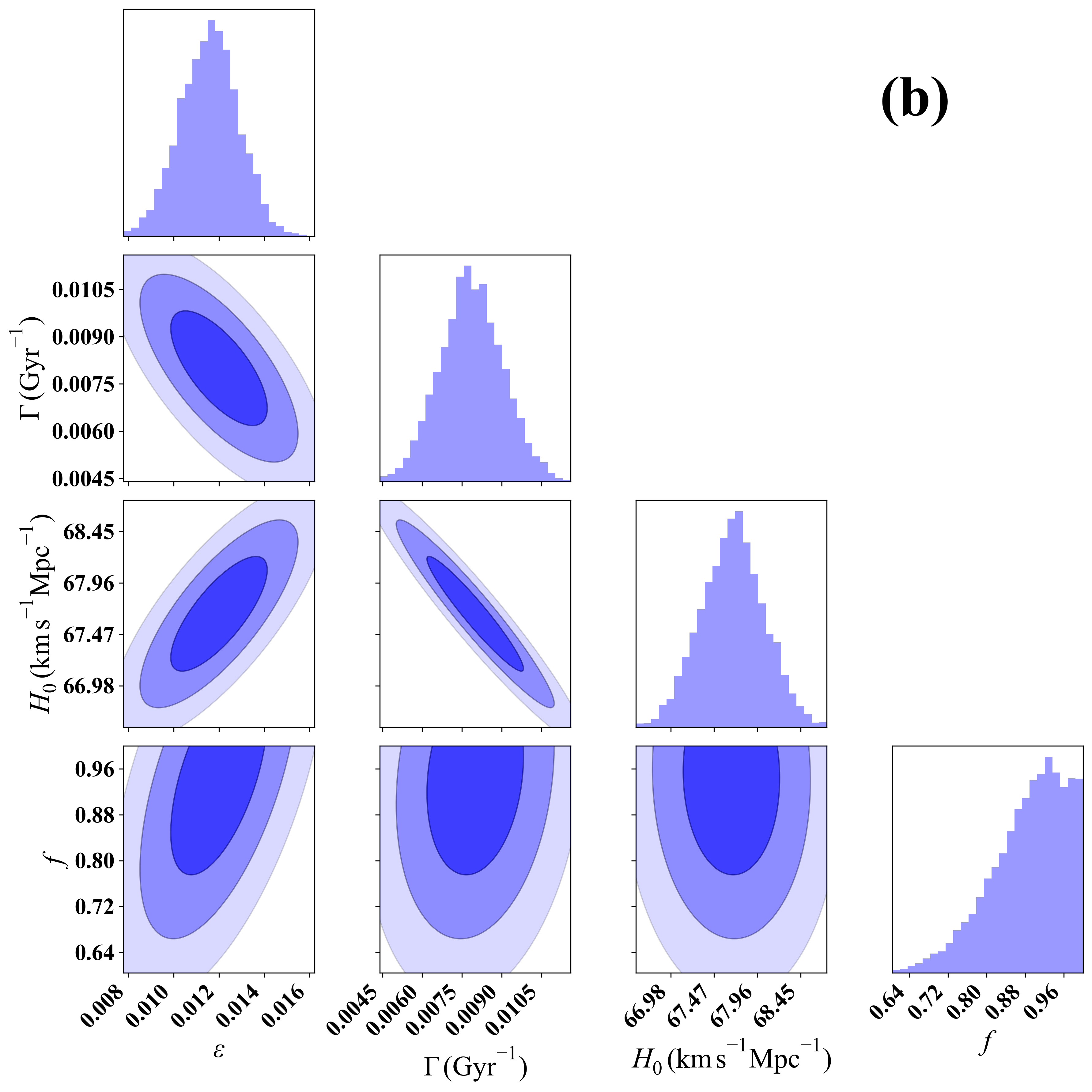}
    \includegraphics[width=0.33\textwidth]{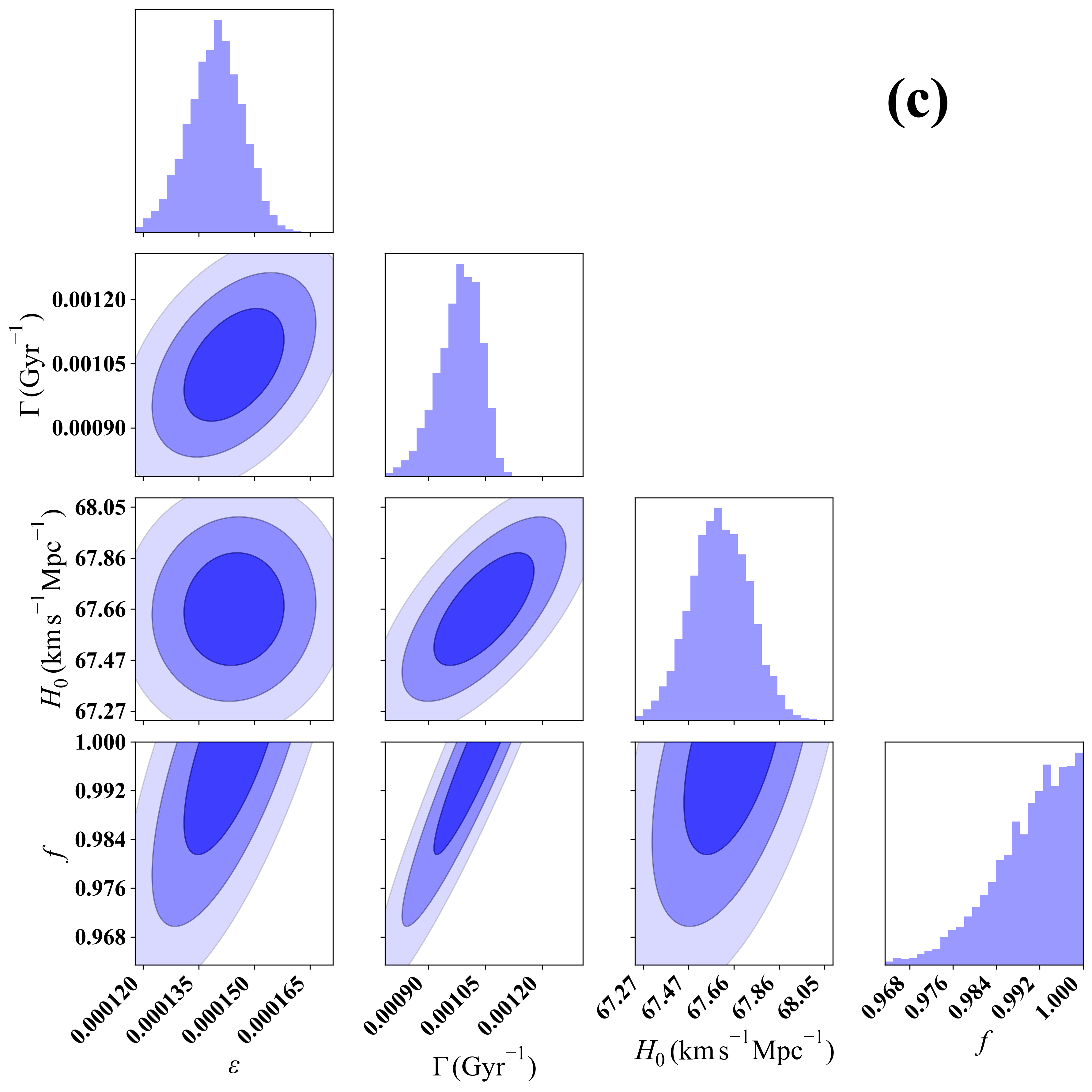}
    \includegraphics[width=0.33\textwidth]{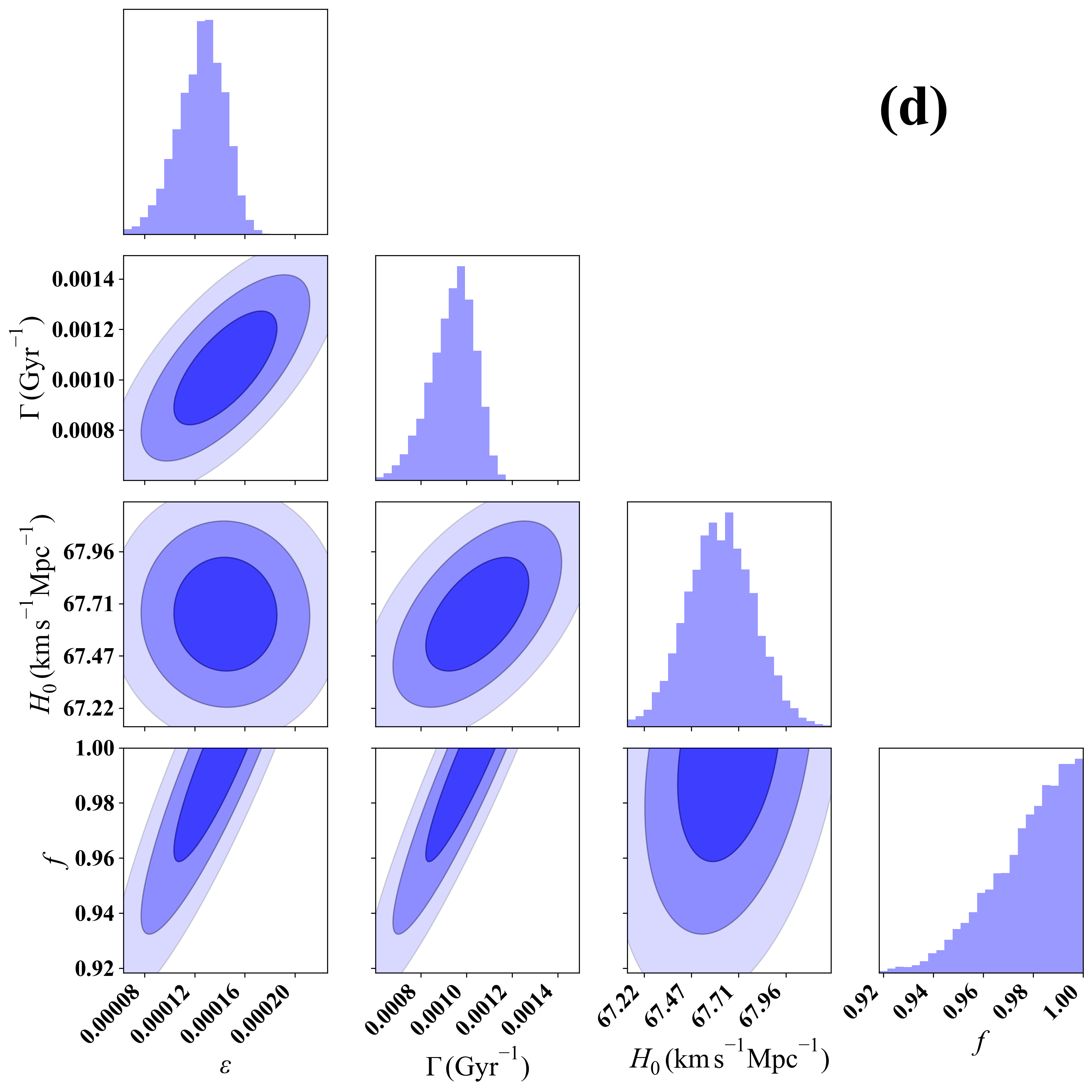}
        \caption{(a) Error projections at the $68 \%$ and $95 \%$ and $99\%$ confidence levels for the parameters ($H_o$, $\Gamma$, $\epsilon$, $f$)=$(67.73 {\rm Km/s/Mpc},0.008 {\rm Gyr^{-1}},0.012,1.0)$. Foregrounds are absent in this analysis
        (b) Error projections at the $68 \%$ and $95 \%$ and $99\%$ confidence levels for the parameters ($H_o$, $\Gamma$, $\epsilon$, $f$)= $(67.73 {\rm Km/s/Mpc},0.008 {\rm Gyr^{-1}},0.012,1.0)$ when the modes contaminated by foregrounds are removed from the analysis.
        (c) Error projections at the $68 \%$ and $95 \%$ and $99\%$ confidence levels for the parameters ($H_o$, $\Gamma$, $\epsilon$, $f$)=$(67.73 {\rm Km/s/Mpc},1.04 \times10^{-3} {\rm Gyr^{-1}},1.14\times10^{-4},1.0)$. Foregrounds are absent in this analysis
        (d) Error projections at the $68 \%$ and $95 \%$ and $99\%$ confidence levels for the parameters ($H_o$, $\Gamma$, $\epsilon$, $f$)= $(67.73 {\rm Km/s/Mpc},1.04 \times10^{-3} {\rm Gyr^{-1}},1.14\times10^{-4},1.0)$ when the modes contaminated by foregrounds are removed from the analysis.}
    \label{fig:contour}
\end{figure*}

 We first consider the case when a perfect foreground cleaning is achieved.
If we assume a fiducial cosmology,  then errors on DDM  parameters can be obtained using the Fisher matrix.
We choose fiducial parameter values $(\epsilon, \Gamma, H_0, f) = ( ~0.012,  ~0.008 {\rm Gyr^{-1}}, ~67.73 {\rm Km/s/Mpc}, ~1.0)$ as the best fit values  from other probes (like the BOSS Lyman-alpha studies \cite{2023lea}).
Fig. \ref{fig:contour}(a) shows the $68 \%$ and $95 \%$ and $99\%$ error contours. The $1-\sigma$ errors are given in Table \ref{tab:1}(A) . 
 These errors are consistent with the literature \cite{2023lea}.
We have marginalized over the well-constrained parameter $\Omega_{mo}$. The correlation between the parameters is on expected lines with an increase in $\epsilon$ and $\Gamma$, implying greater suppression of power and over a greater range of scales, respectively.  We find that for the given fiducial cosmology and a futuristic SKA-like observation, we can measure $\epsilon$  and $\Gamma$ at $3.37\%$ and $6.86 \%$ respectively.

We note here that the dark energy sector has been kept entirely untouched in our analysis, and we have assumed a constant dark energy equation of state $w =-1$. Any dynamical dark energy equation of state would affect both the background evolution and the structure formation. The amplitude of the power spectrum $\mathcal{C}(z)$ is sensitive to the IGM astrophysics and astrophysical parameters like the neutral fraction, or the large-scale \nh~ bias, have strong degeneracy with $H_o$ and $\sigma_8$. 
We, therefore, focus only on the parameters of the DDM model.

While 21-cm intensity mapping is a promising probe of cosmology beyond the $\Lambda$CDM  model, such experiments are plagued with the problem of foregrounds originating from sources like galactic Synchrotron radiation, free-free emission, or extragalactic radio point sources \cite{di2002radio,shaver1999can,2011MNRAS.418.2584G}. These foregrounds are $\approx 5$ orders of magnitude larger \cite{mcquinn2006cosmological,santos2005multifrequency} than the signal and present a formidable challenge towards detecting the 21-cm signal. 
While the issue of foregrounds is less challenging for the post-reionization signal as compared to the epoch of reionization, significant foreground cleaning is required before the cosmological signal can be explored.  
The difference in the spectral properties of the foregrounds from that of the  21-cm signal is often exploited to  remove the
foregrounds from observed visibility data or from the multifrequency angular power spectrum\citep{paciga2011gmrt, datta2010bright, chapman2012foreground, mertens2018statistical, trott2022multi, datta2007multifrequency, 2011MNRAS.418.2584G, 2023MNRAS.525.3439E}.
For a spectrally smooth foreground, the small $k_{\parallel}$ modes are affected.
Foreground from sources in and around the field of view can be removed by avoiding the modes in the "foreground wedge" while estimating the 21-cm power spectrum \cite{pober2013opening, pober2014next, liu2014epoch, dillon2015empirical, pal2021demonstrating}.
The equation of the foreground wedge boundary is given by \cite{Seo_2016}
\begin{equation}
k_{\parallel} = k_{\perp}\frac{\sin\theta_{\mathrm{FoV}}\,E(z)}{(1+z)}\int_0^z \frac{dz'}{E(z')}
\end{equation}
where $\theta_{\mathrm FoV}$ is the field of view of the telescope.
 We must note that the wedge is not a fundamental
effect, and in fact, the foregrounds in this region may be removed by perfect baseline-to-baseline calibration. However, raw systematics dominate much strongly over the cosmological signal in the wedge. 
It is thus safe to remove the modes in the wedge for parameter estimation.
Fig \ref{fig:wedge} shows the modes to be removed (grey wedge zone) at the three observing frequencies.

Fig. \ref{fig:contour}(b) shows the $68 \%$ and $95 \%$ and $99\%$ error contours, when the modes in the wedge are removed from the Fisher analysis. The error estimates are severely degraded due to the loss of many modes. In this case $\epsilon$ and $\Gamma$ are measured at $9.91 \%$
and $12.03\%$ respectively.

Table \ref{tab:1}(B) shows the degraded error estimates when we incorporate the foreground wedge in our analysis.

  The bias $b_T(k,z)$ plays a central role in the modeling of the post-reionization 21-cm signal. A full N-body simulation of the decaying dark matter scenario is required to model the bias. That is outside our present scope. While we have adopted the bias obtained from cold dark matter simulations, we note that the imprint of the power suppression on small scales on the bias is not incorporated. As discussed earlier, this leads to an underestimation of the bias and hence the 21-cm signal.  If we assume the bias parameters to be a part of the Fisher analysis and subsequently marginalize over them, then the errors on the crucial DDM parameters are highly degraded. Table \ref{tab:1}(C) shows the large uncertainties in the parameters.
This is expected, since the bias fixes the overall amplitude of the power spectrum. Apart from that, the bias also determines the anisotropies in redshift space. The forecast under identical assumptions for a second, literature-anchored benchmark $(\epsilon,\Gamma)=\big(1.14\times10^{-4},\;1.04\times10^{-3}~\mathrm{Gyr}^{-1}\big)$ yields projected fractional uncertainties of $6.16\%$ on $\epsilon$ and $8.29\%$ on $\Gamma$; restricting to the foreground-free window increases these to $18.77\%$ and $14.22\%$, respectively.Fig. \ref{fig:contour}(c) and  \ref{fig:contour}(d) shows the $68 \%$ and $95 \%$ and $99\%$ error contours for the cases when foreground wedge is absent and when foreground wedge is present respectively. The $1-\sigma$ errors are given in Table \ref{tab:1}(D) and \ref{tab:1}(E).Further,when the bias is marginalized over, the errors are highly degraded.

Apart from foregrounds, RFI's pose a critical problem in detecting the signal.
Further, bandpass calibration errors introduce spurious spectral structure
into the foreground signal. 
This difficulty has led to many proposals for precise bandpass calibration \cite{mitchell2008real,  kazemi2011radio,  sullivan2012fast, kazemi2013robust, dillon2020redundant, kern2020absolute, byrne2021unified, sims2022bayesian, ewall2022precision}.
These major observational issues would degrade the signal, and a more complete, realistic analysis invoking some of these issues is necessary.

 We conclude by noting that the post-reionization 21-cm intensity mapping has the potential to probe 2-body decaying dark matter models and constrain model parameters. This holds the promise for a better understanding of the dark matter sector and sheds valuable light on possible dark matter candidates.  However, we also remind ourselves that several challenging observational and modeling issues need to be dealt with before this is achieved.

\clearpage
\bibliography{mybib}

\end{document}